\documentclass[pra,aps]{revtex4-2}
\pdfoutput=1
\usepackage{amsmath,amssymb}
\usepackage{graphicx}% Include figure files
\usepackage{color}
\usepackage{dcolumn}% Align table columns on decimal point
\usepackage{bm}% bold math
\usepackage{array}
\usepackage{multirow}
\usepackage{hyperref}
\hypersetup{colorlinks=true,linkcolor=blue,citecolor=blue}

\begin{document}

\title{Quantum state transfer of superposed multi-photon states via phonon-induced dynamic resonance in an optomechanical system}

\author{Xuanloc Leu}
\author{Xuan-Hoai Thi Nguyen}
\author{Jinhyoung Lee}
\email{hyoung@hanyang.ac.kr}
\affiliation{Department of Physics, Hanyang University, Seoul 04763, Korea}

\begin{abstract}
We propose a method to transfer macroscopically superposed states between two optical cavities mediated by a mechanical oscillator, which works in a nonlinear regime of optomechanical interaction. The scheme relies on phonon-induced dynamic resonance, where the motion of the mechanical oscillator dynamically sets on/off the resonance between the two cavities. It assumes the high-amplitude limit of the oscillator, weak coupling between optical cavities, and the adiabatic approximation. We show that, under these conditions, various multi-photon quantum states, especially Schr{\"o}dinger cat states, can be faithfully transferred in a deterministic process. Our approach thus suggests the possibility of transferring large-photon-number quantum states, beyond the few-photon regime typically accessible with linearized optomechanical interactions.
\end{abstract}

%%%%%%%%%%%%%%%%%%%%%%%%%%  body  %%%%%%%%%%%%%%%%%%%%%%%%%%

\maketitle
\newcommand{\bra}[1]{\left<#1\right|}
\newcommand{\ket}[1]{\left|#1\right>}
\newcommand{\abs}[1]{\left|#1\right|}
\newcommand{\expt}[1]{\left<#1\right>}
\newcommand{\braket}[2]{\left<{#1}|{#2}\right>}
\newcommand{\commt}[2]{\left[{#1},{#2}\right]}

\newcommand{\I}{\mathsf{i}}
\newcommand{\E}{\mathsf{e}}

%add subtitle for figures
\newsavebox{\panelbox}

\newcommand{\toppanel}[2]{%
    \sbox{\panelbox}{#1}%
    \usebox{\panelbox}%
    \llap{%
        \raisebox{\dimexpr\ht\panelbox-0.0em\relax}{%
            \makebox[\wd\panelbox][l]{%
                \hspace{-0.3em}\small (#2)}}}%
}

%============Section 1====================================================================

\section{Introduction}

Quantum state transfer (QST), the faithful transmission of quantum states between physical systems or locations, is a central task in quantum information processing and communication~\cite{QIP1,QIP2,QIP3,QIP4,QIP5}. Representative approaches include quantum teleportation~\cite{QT1,QT2,QT3,QT4,QT5}, double-swap protocols~\cite{Swap}, and adiabatic passage through dark states~\cite{APP,STIRAP1,STIRAP2,STIRAP3}. These methods have been widely studied in atomic~\cite{atom1a,atom1b,atom1c,atom1d,atom2a,atom2b,atom3a,atom3b,atom3c,atom4,atom5a,atom5b,atom5c,atom6a,atom6b,atom7a,atom7b}, spin~\cite{spin1a,spin1b,spin2a,spin2b,spin2c,spin2d,spin3}, ionic~\cite{ion1a,ion1b,ion2a,ion2b,ion2c,ion3}, superconducting-circuit~\cite{scc1a,scc1b,scc1c,scc1d,scc1e,scc1f,scc1g,scc1h,scc2,scc3}, solid-state~\cite{ss1,ss2a,ss2b,ss2c}, and optical systems~\cite{opt1,opt2a,opt2b}.

Optomechanical systems (OMSs), in which optical and mechanical modes interact through radiation pressure~\cite{MTK}, provide a versatile platform for QST. Advances in mechanical cooling and strong optomechanical coupling~\cite{SCExp1,SCExp2,SCExp3,SCExp4,SCExp5} have enabled optical--mechanical state transfer~\cite{CMTheo1,CMTheo2,CMTheo3,CMTheo4,CMTheo5a,CMTheo5b,CMTheo6,CMTheo7,CMExp2,SCExp6,CMExp1,CMExp3}. In three-mode OMSs, both double-swap and adiabatic protocols have been proposed for QST between optical modes of different frequencies~\cite{CMCTheo1,CMCTheo2,CMCTheo3,CMCTheo4,CMCTheo5} and demonstrated experimentally through coherent wavelength conversion~\cite{CMCExp1,CMCExp2,CMCExp3,CMCExp4}. Related developments include hybrid protocols~\cite{HS1,HS2,HS3}, shortcuts to adiabaticity~\cite{StoA1,StoA3}, transfer between mechanical resonators~\cite{MCM1,MCM2,MCM3,MCM4}, distant OMSs~\cite{DisOMS1,DisOMS2,DisOMS3}, and hybrid atom--optomechanical systems~\cite{hybrid1a,hybrid1b,hybrid2,hybrid3,hybrid4}. Progress in optomechanical entanglement control~\cite{Ent} also opens a route toward quantum teleportation in OMSs~\cite{Tel1,Tel2,Tel3,Tel4,Tel5,Tel6}.

Most existing optomechanical QST schemes rely on a linearized beam-splitter interaction~\cite{MTK}, limiting the amplitude of the quantum states that can be transferred. The limitation arises from a built-in trade-off: A strong pumping field is required to enhance the coherent coupling well beyond the cavity damping rate~\cite{CMCTheo1,CMCTheo3,CMCTheo4}, yet this regime of the pumping field is precisely where the linear approximation breaks down~\cite{CMTheo5a,CMTheo5b}. Meanwhile, the effective coupling must remain sufficiently below the mechanical frequency to suppress unwanted counter-rotating or Stokes processes. These competing requirements restrict the coherent coupling rate to be on the order of the cavity damping rate, which limits the average photon number of transferable quantum states to just a few~\cite{CMExp1,SCExp6,CMExp3}.

While much attention has focused on microwave–optical conversion as a key application of OMS, mechanically mediated optical--optical transfer has also been demonstrated for wavelength conversion and optical storage~\cite{CMCExp1,add2} and studied as a tunable photon--photon interface~\cite{DisOMS1}. Such transfer is advantageous for realizing a reconfigurable interconnect between different on-chip optical modes, complementing static waveguide links, and is naturally compatible with emerging integrated photonic platforms~\cite{add4,add5}.

In this work, we propose a model to transfer multi-photon quantum states between two coupled optical cavities via a mechanical oscillator, operating in a nonlinear regime. We consider a three-mode OMS in which two cavities optomechanically couple to a movable in-between mirror~\cite{Law}, while also interacting each other. Rather than relying on a linearized beam-splitter interaction, we exploit the dynamic resonance transfer~\cite{JLim}: mechanical motion dynamically brings the initially detuned cavity modes into resonance, by which a quantum state is transferred between the cavity modes of different wavelengths without necessitating linearization. In this sense, the phonon-induced dynamic resonance is extended from single-excitation energy transport in molecular networks~\cite{JLim} to the coherent transfer of superposed multi-photon bosonic states in the full cavity Hilbert space, where preservation of phase coherence is essential.

The mechanism requires a high-amplitude mechanical oscillator, weak optical inter-cavity coupling, and adiabatic approximation. In the high-amplitude regime, coherent mechanical motion dominates quantum and thermal fluctuations, allowing a semiclassical treatment of the oscillator, while the weak-coupling and adiabatic conditions enable the optical state to follow the instantaneous cavity eigenmodes as the mechanically induced resonance is crossed. We show that this nonlinear mechanism can support high-fidelity transfer of large-amplitude Fock, displaced-squeezed, and Schr\"odinger-cat states, which are central to quantum optics and quantum information processing~\cite{Knight,JLee,SCS1,SCS2,SCS3}.

%============Section 2====================================================================
\section{System and its dynamics}\label{sec2}

%--------fig 1: Model's schemetic--------------------------
\begin{figure}[t]
\begin{center}
\includegraphics[height=0.25\textwidth]{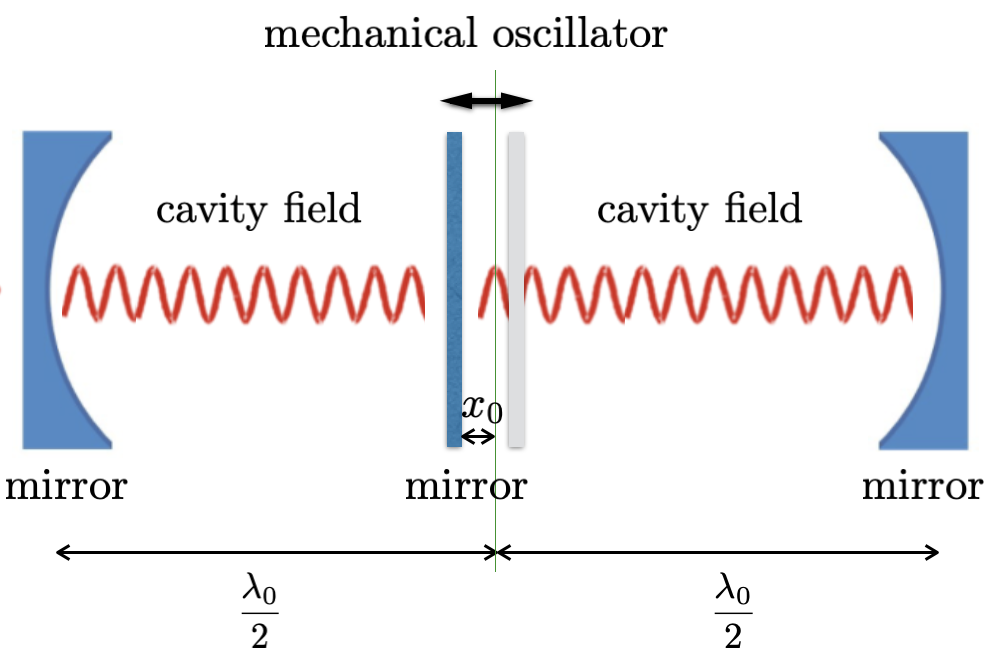}
\end{center}
\caption{Schematic of a representative three-mode optomechanical system. Two cavities share a vibrating in-between mirror of mass $m$ and frequency $\omega_m$. Its initial displacement $x_0\ll\lambda_0$ from the center gives cavity frequencies $\omega_{1,2}=2\pi c/\lambda_{1,2}$ with $\Delta\omega=\omega_1-\omega_2=2\delta\omega$. The corresponding dimensionless initial displacement is defined as $b_0=x_0/2x_{\rm ZPF}$. The vibrating mirror is slightly transparent, coupling the two cavities with strength $g$, and interacts optomechanically with both cavity modes with equal couplings $\kappa_1=\kappa_2=\kappa_0$.}
\label{fig:OMS4}
\end{figure}
%--------fig 1: end-------------------

We consider an optomechanical model consisting of two cavities sharing an in-between mirror of mass $m$ and frequency $\omega_m$, as depicted in Fig.~\ref{fig:OMS4}. Rather than being centered, which would give the two cavities the same fundamental frequency $\omega_0=2\pi c/\lambda_0$, the mirror is initially located a distance $x_0>0$ to the left of the center, with $x_0\ll\lambda_0$. The fundamental frequencies of the left and right cavities, denoted by $\omega_1$ and $\omega_2$, respectively, therefore have a mismatch $\Delta\omega=\omega_1-\omega_2$, where $\omega_{1(2)}=2\pi c/\lambda_{1(2)}$, with $\lambda_1=\lambda_0-2x_0$ and $\lambda_2=\lambda_0+2x_0$.
The vibrating mirror, called a mechanical oscillator (or simply an oscillator), is assumed to be slightly transparent so as to couple the two cavities with a coupling constant $g>0$. 

In a frame rotating at $\omega_0$, the effective Hamiltonian is given by ($\hbar=1$)~\cite{Law}
%_______Eq: H_eff______________
\begin{equation}\label{H4}
\hat{H}_{{\rm eff}} =  g (\hat{a}_{1}^\dag \hat{a}_{2} +  \hat{a}_{1}\hat{a}_{2}^\dag ) + 2 \kappa_0 b_0 (1 - \frac{\hat{b} + \hat{b}^\dag}{2 b_{0}}) \Delta \hat{n} + \omega_m \hat{b}^\dag \hat{b}.
\end{equation}
%_______Eq: end_____________
Here, $\hat{a}_{1(2)}$ and $\hat{b}$ are the annihilation operators of the left (right) cavity and the oscillator, respectively, and $\Delta\hat{n}=\hat{n}_1-\hat{n}_2$ with the number operators $\hat{n}_j=\hat{a}^\dagger_j\hat{a}_j$ for $j=1,2$. The optomechanical coupling constant $\kappa_0 = 2x_{\rm ZPF}\, \omega_{0} / \lambda_0$ and the dimensionless shift $b_0 = x_0/2x_{\rm ZPF}$ with $x_{\rm ZPF}=(\hbar/2m\omega_m)^{1/2}$ being the mechanical zero-point fluctuation amplitude~\cite{MTK}. The first term in equation~(\ref{H4}) describes the optical coupling between cavities, the second term the optomechanical interaction between the oscillator and the cavities, and the third term the free energy of the oscillator. We neglect position-dependent hopping, $(\hat{b}+\hat{b}^\dag)(\hat{a}_1\hat{a}_2^\dag + \hat{a}_1^\dag\hat{a}_2)$, because the mirror motion is modeled as a rigid translation. Higher-order optomechanical interactions, such as $(\hat{b}+\hat{b}^\dag)^2 \hat{a}_j^\dag\hat{a}_j$, are also neglected since the mirror equilibrium position is biased away from the symmetry point~\cite{mem1, mem2}.

Taking into account damping and noise, the Langevin equations for the system operators can be written as~\cite{GZ04}
%_______Eq: Langevin eqs______________
\begin{equation}\label{dnm}
i\dot{{\bf v}}(t)
={\bf M}(t){\bf v}(t)
+{\bf f}_{\rm rp}(t)
+i{\bf f}_{\rm in}(t),
\end{equation}
%_______Eq: end_____________
where
${\bf v}(t)=\big(\hat{a}_1(t),\hat{a}_2(t),\hat{b}(t)\big)^T$
is the vector of annihilation operators and
$\dot{{\bf v}}=d{\bf v}/dt$.
The radiation-pressure backaction on the oscillator is represented by
${\bf f}_{\rm rp}(t)=\big(0,0,-\kappa_0\Delta\hat{n}(t)\big)^T$,
whereas
$
{\bf f}_{\rm in}(t)
=
\left(
\sqrt{2\gamma_1}\hat{s}_1^{\rm in}(t),
\sqrt{2\gamma_2}\hat{s}_2^{\rm in}(t),
\sqrt{2\gamma_m}\hat{s}_m^{\rm in}(t)
\right)^T
$
describes the input noise associated with the three dissipative channels.
Here, $\gamma_{1,2}$ and $\gamma_m$ are the decay rates of cavities 1 and 2 and the mechanical oscillator, respectively. We assume independent reservoirs satisfying
$\langle\hat{s}^{\rm in}_j(t)\hat{s}^{\rm in\dagger}_j(t')\rangle
=(n_j^{\rm th}+1)\delta(t-t')$ and $\langle \hat{s}^{\rm in\dagger}_j(t)\hat{s}^{\rm in}_j(t') \rangle
=n_j^{\rm th}\delta(t-t')$,
where $n_j^{\rm th}$ ($j=1,2,m$) denotes the corresponding thermal occupation.
The dynamical matrix ${\bf M}(t)$ is
%_______Eq: dynamic matrix M_____________
\begin{equation}\label{QDE}
\mathbf{M}(t)=\left(
\begin{array}{ccc}
\hat{\omega}(t)-i\gamma_1 & g & 0 \\
g & -\hat{\omega}(t)-i\gamma_2 & 0 \\
0 & 0 & \omega_m-i\gamma_m
\end{array}
\right),
\end{equation}
where 
\begin{equation}
\hat{\omega}(t):=2\kappa_0b_0\left[1-\big(\hat{b}(t)+\hat{b}^\dag(t)\big)/2b_0\right]
\end{equation}
is the mechanically modulated cavity-detuning operator.

It is worth noting that the dynamical equation in equation~(\ref{dnm}) is nonlinear. Nevertheless, if the oscillator is treated classically and oscillates slowly enough ($\omega_m \ll g$), the dynamics simplify to a beam-splitter-type interaction between two cavities with time-dependent frequencies. This interaction becomes resonant when the frequencies of the two become mutually equal depending on the oscillator's position. This phenomenon, known as phonon-induced dynamic resonance~\cite{JLim}, is discussed in more detail in the next section.

%============Section 3====================================================================
\section{Phonon-induced dynamic resonance in adiabatic process}\label{sec3}

In this section, we show how our model enables the manifestation of phonon-induced dynamic resonance between two cavities. We first provide an intuitive classical picture and then analyze the quantum dynamics with numerical illustrations. Meanwhile, we examine the conditions required for dynamic resonance with high transmission and establish a self-consistent parameter regime that simultaneously satisfies all these conditions.

%--------fig added: illustration of PIDR--------------------------
\begin{figure}[t]
\begin{center}
\includegraphics[height=0.1\textwidth]{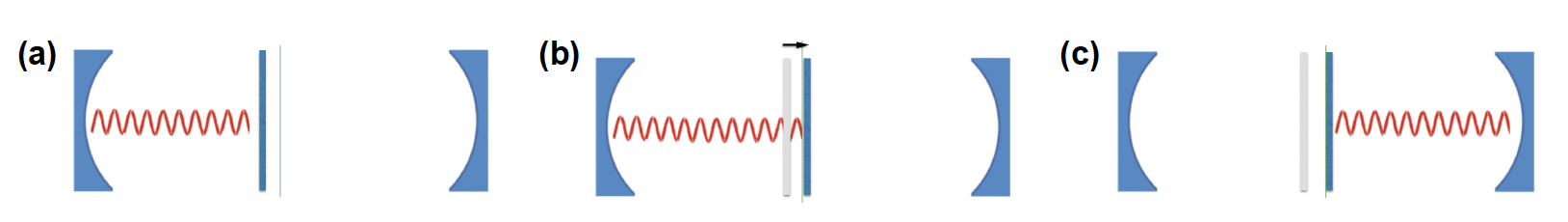}
\end{center}
\caption{Intuitive illustration of quantum state transfer via phonon-induced dynamic resonance. (a) Photons initially occupy the far-detuned left cavity and push the in-between mirror to the right. (b) As the mirror approaches the center, the cavity frequencies become resonant, enabling photon tunneling. (c) After transfer to the right cavity, radiation pressure drives the mirror back and restores the detuning. This self-induced resonance modulation enables deterministic state transfer without external driving.}
\label{fig:add}
\end{figure}
%--------fig added: end---------------------------------------------

We assume a weak coupling between optical cavities, characterized by $g\ll \delta\omega=(\omega_1-\omega_2)/2\approx 2\kappa_0b_0$, so that they are initially in far-off resonance. In contrast, under strong optical coupling, direct hopping between the cavities dominates, leading to a conventional quantum state exchange~\cite{CROW}. We also assume that
the mechanical motion is slow compared with the optical dynamics,
$\omega_m\ll g$. 
Let us suppose that the initial quantum state has been prepared in the left cavity and the oscillator starts leaving  its initial equilibrium position, $\langle\hat{b}(0)\rangle=0$, at $t=0$. The radiation pressure by the left cavity pushes the oscillator toward a new equilibrium position at the right. The oscillator then approaches the center of the two cavities, depending on the excited energy in the left cavity, so that the cavities become resonant and exchange photons. The resonance condition is temporally satisfied with the oscillator's position, so-called a phonon-induced dynamic resonance~\cite{JLim}. Figure~\ref{fig:add} illustrates how the mirror motion dynamically tunes the cavities into and out of resonance, enabling photon transfer from the left to the right cavity. 

The oscillator's equilibrium position of average $b_{\rm eq}= \kappa_0\langle\Delta \hat{n}\rangle/\omega_m$ is determined by the radiation pressures or the photon numbers of the cavity fields. 
The initial photon-number average in the left cavity, denoted as $\bar{n}$, needs to be so large that the dynamic resonance condition can be fulfilled with $2b_{\rm eq}>b_0$, or equivalently, $\bar{n}>n_{\rm thr}$, where the threshold number of photons $n_{\rm thr}=\omega_mb_0/2\kappa_0$. In this regime of dynamic resonance and the time duration much shorter than $2\pi/\gamma_m$ (i.e., $\gamma_m\ll\omega_m$), one may neglect the quantum fluctuation $\delta\hat{b}$ of the oscillator if the following condition is satisfied~\cite{Lud}
%_______Eq: high amplitude_____________
\begin{equation}\label{HAC}
 \sqrt{ \langle \delta \hat{b}^\dag \delta \hat{b}  \rangle} \ll \abs{b-b_{{\rm eq}}}\simeq b_{{\rm eq}}.
\end{equation}
Here we expand $\hat{b}=b+\delta\hat{b}$, where the average amplitude $b=\langle\hat{b}\rangle$ is the solution to $i\dot{b}=\omega_m(b-b_{\rm eq})$. The high amplitude limit in equation~(\ref{HAC}), as assumed throughout this paper, is that the average amplitude from the equilibrium, $|b-b_{\rm eq}|$, is much larger than the quantum fluctuation~\cite{Lud}. In terms of the number fluctuation of cavities, the condition in equation~(\ref{HAC}) is reformulated as
%_______Eq: high amplitude: cavity_____________
\begin{equation}
\label{HAC_cav}
\sqrt{\langle(\Delta\hat{n}-\langle\Delta\hat{n}\rangle)^2\rangle}\ll|\langle\Delta\hat{n}\rangle|.
\end{equation}
In this limit, the frequency operator $\hat{\omega}(t)$ in equation~(\ref{QDE}), a function of $\hat{b}$, is reduced to the scalar function $\omega(t)$ of the average $b$, i.e., $\hat{\omega}(t)$ is replaced by $\omega(t)=\delta\omega(1-b_r(t)/b_0)$, where $b_r$ is the real part of $b$. In the oscillator's equation, we replace the operators by their averages, i.e., $\hat{b}(t)$ by $b(t)$ and $\Delta\hat{n}(t)$ by $\langle\Delta\hat{n}(t)\rangle$. We also remove the noise operator $\sqrt{2\gamma_m}\hat{s}^{\rm in}_m(t)$ by assuming the thermal phonon number $n^{\rm th}_m\ll |b-b_{\rm eq}|^2$. The dynamic equation~(\ref{dnm}) becomes, divided into two types,
%_______Eq: Langevin eqs: separated_____________
\begin{eqnarray}
 i \dot{{\bf a}}(t) &=& {\bf M}_a(t) {\bf a}(t)+i{\bf f}^{\rm in}_a(t), \label{QDE11}\\
 i \dot{b}(t) &=& (\omega_m-i\gamma_m)b(t) - \kappa_{0}  \langle \Delta \hat{n}(t) \rangle, \label{QDE12}
\end{eqnarray}
where ${\bf a}(t)=\big(\hat{a}_1(t),\hat{a}_2(t)\big)^T$ and ${\bf f}^{\rm in}_a(t)=\left(2\sqrt{\gamma_1}\hat{s}^{\rm in}_1(t),\sqrt{2\gamma_2}\hat{s}^{\rm in}_2(t)\right)^T$ are vectors for the cavities, and
%_______Eq: dynamic matrix for cavities_____________
\begin{equation}\label{DM}
{\bf M}_a(t)=\left(
\begin{array}{ccc}
\omega(t)-i\gamma_1 & g\\
g & -\omega(t)-i\gamma_2
\end{array}
\right).
\end{equation}

%------fig: eigen frequencies---------------
\begin{figure}[t]
\begin{center}
\includegraphics[height=0.2\textwidth]{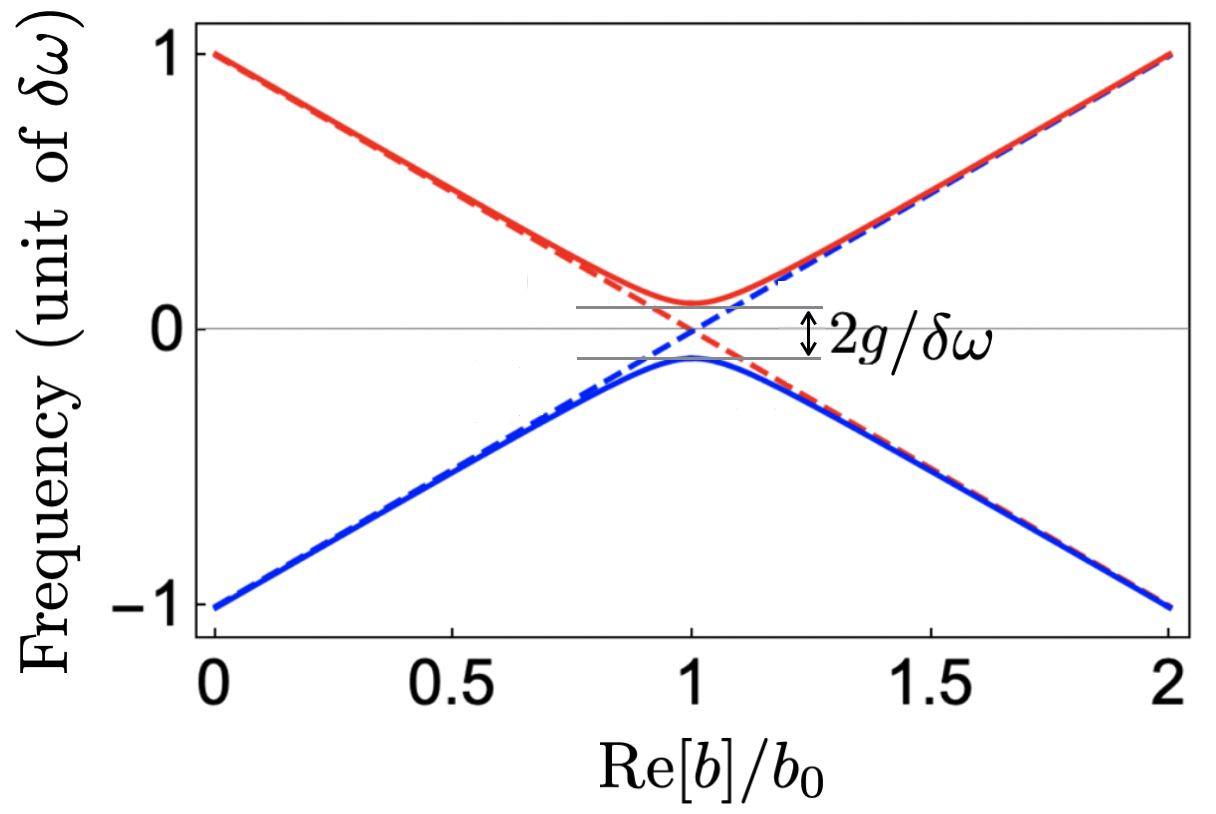}
\end{center}
\caption{Diagram of optical frequencies (in unit of $\delta\omega$) in terms of the normalized mechanical displacement ${\rm Re}[b]/b_0$. For $g=0$, the bare frequencies (dashed) $\omega_{1,2}(b)=\pm\omega(b)$ cross linearly at ${\rm Re}[b]/b_0=1$. For $g>0$, the eigen-mode frequencies (solid) $\pm\epsilon(b)=\pm\sqrt{\omega^2(b)+g^2}$ show an avoided crossing with a gap $2g$. The near touching of the two branches, attainable for small $g/\delta\omega$, characterizes the adiabatic process.}
\label{fig:EV}
\end{figure}
%------fig: end----------------------------------

Figure~(\ref{fig:EV}) presents the instantaneous optical eigenfrequencies, given by
the matrix ${\bf M}_a(t)$, as functions of $b_r/b_0$, where
$b_r=\mathrm{Re}[b]$. It shows a typical feature of the adiabatic process. For zero optical coupling of $g=0$, the width of the left (right) cavity is linearly increased (decreased) when the oscillator moves to the right. As a result, the frequency of cavity mode $\hat{a}_1$ ($\hat{a}_2$) is linearly decreased (increased), leading to a frequency crossing at $b_r/b_0=1$. When $g>0$, a gap emerges between the two eigen-mode frequencies, characterized by the order of $g/\delta\omega$. It's essential for this order to be small enough
 that the two frequencies nearly touch each other, supporting dynamically induced transmission. 

For the cavity dynamics to adiabatically follow these instantaneous eigenmodes, we require 
\begin{equation}
\nu\approx \frac{1}{8}\frac{\bar n}{n_{\rm thr}}
\frac{\omega_m\delta\omega}{g^2}\ll1,
\end{equation}
as derived in Appendix~\ref{apdA}.
Combining this condition with the weak optical coupling
$g/\delta\omega=\varepsilon\ll1$ and the dynamic-resonance condition
$\bar n/n_{\rm thr}=r>1$, we obtain
%_______Eq: Adiabatic cond_______________________
\begin{equation}\label{AdbCon}
\frac{\kappa_0}{\omega_m}\simeq\frac{r}{4\varepsilon \sqrt{2\bar n \nu}}>\frac{1}{4\varepsilon \sqrt{2\bar n \nu}},
\end{equation}
which sets a lower bound on the optomechanical coupling $\kappa_0$. The condition for neglecting the thermal fluctuation of the oscillator, $n_{m}^{\rm th} \ll |b_{\rm eq}|^2$, can
then be written as
\begin{equation}\label{negl_thermal}
n_{m}^{\rm th} \approx \frac{k_B T}{\hbar \omega_m} \ll \frac{r^2 \bar n}{32 \varepsilon^2 \nu}.
\end{equation}

To estimate the transmission efficiency and choose the system parameters, we use the ideal analytical solution derived in Appendix~\ref{apdA2}, which gives the averaged transmittivity after the oscillator passes the resonant position as
\begin{equation}\label{Tave}
\overline{|T_{21}|}\approx1-\frac{\varepsilon^2}{4},
\end{equation}
where $\varepsilon=g/\delta\omega\ll1$. Equation~(\ref{Tave}) shows that, once the adiabatic condition is satisfied, the ideal transmission efficiency is mainly determined by the ratio $\varepsilon=g/\delta\omega$, consistent with the intuitive picture in figure~\ref{fig:EV}. Throughout this work, we set $\varepsilon=10^{-2}$, giving $\overline{|T_{21}|}\approx 0.9999$.

%------fig: Trajectories of oscillator---------------
\begin{figure}[t]
\begin{center}
\includegraphics[height=0.35\textwidth]{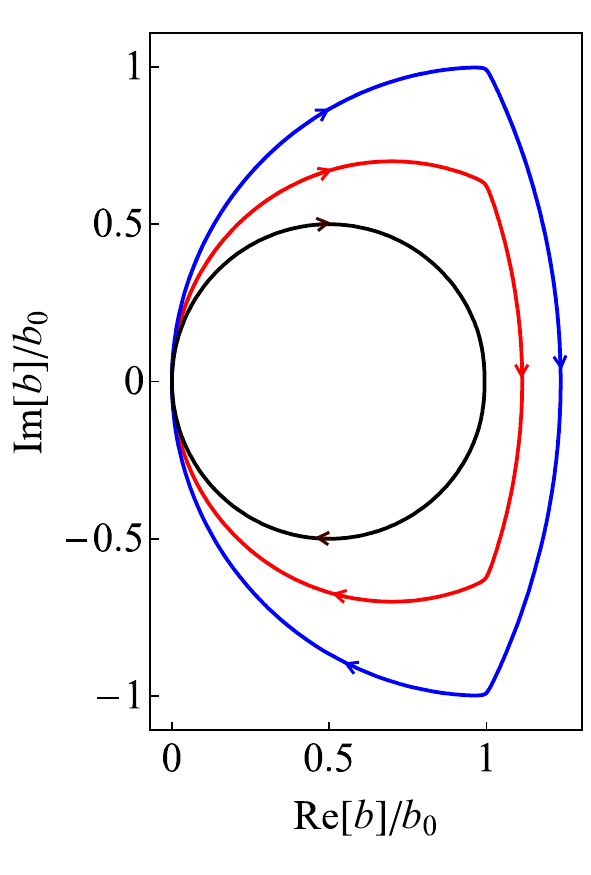}
\end{center}
\caption{Semiclassical trajectories of the mechanical oscillator on phase space when $\bar{n} / n_{{\rm thr}} =$ 1 (black), 1.4 (red), and 2 (blue) for given $\omega_m/g=10^{-3}$ and $g/ \delta \omega = 10^{-2}$. The trajectory starts at the origin and evolves clockwise toward the resonant position ${\rm Re}[b]/b_0=1$ when $\bar{n}>n_{\rm thr}$. At this point, photons are transferred to the right cavity mode (see figure~\ref{fig:PO}), and the oscillator follows the larger circle. Upon returning to the resonant position, the photons transfer back to the left cavity mode (see figure~\ref{fig:PO}), and the oscillator retraces the smaller circle.}
\label{fig:CS}
\end{figure}

%------fig: Photon populations---------------
 \begin{figure}[t]
\begin{center}
\includegraphics[height=0.23\textwidth]{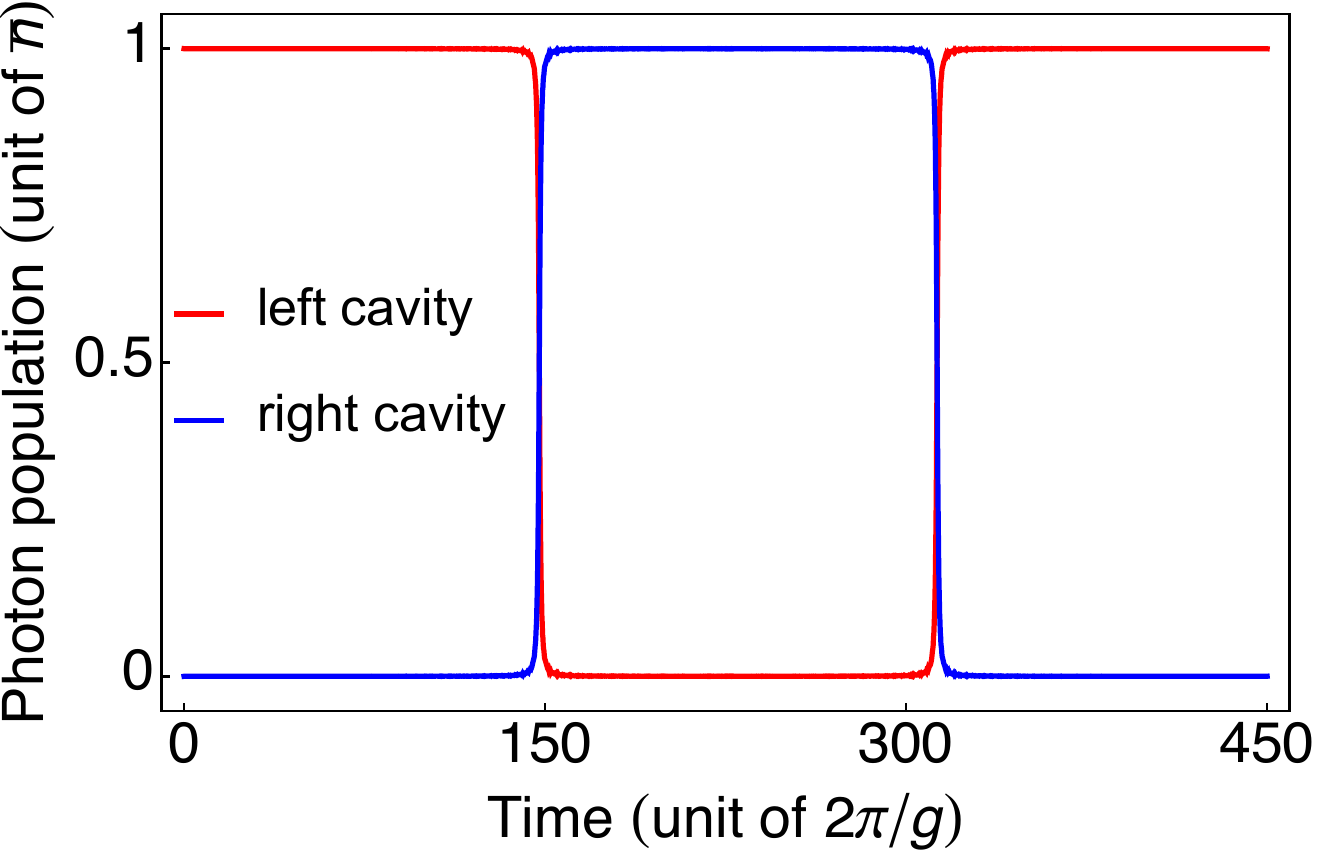}
\end{center}
\caption{Photon populations of two cavities as functions of time, normalized by initial average photon number $\bar{n}$ in the left cavity, for $\bar{n} / n_{{\rm thr}} =5$, $\omega_m/g =  10^{-3}$, and $g/\delta \omega = 10^{-2}$. The populations show complete photon transfer from the left cavity (red) to the right cavity (blue) when the oscillator reaches the resonant position ${\rm Re}[b]/b_0=1$ in figure~\ref{fig:CS}. As the oscillator continues along its phase-space trajectory, the photons are transferred back to the left cavity, producing the periodic exchange dynamics consistent with the loops shown in figure~\ref{fig:CS}.}
\label{fig:PO}
\end{figure}
%------fig: end-------------------------------------

Given $g/\delta\omega=10^{-2}$ and $\omega_m/g=10^{-3}$, we illustrate the phonon-induced dynamic resonant transfer by presenting the semiclassical solution of $b(t)/b_0$ 
on phase space of $({\rm Re}[b],{\rm Im}[b])$ in figure~\ref{fig:CS} and photon populations of cavities in figure~\ref{fig:PO}, for the ideal case of $\gamma_1=\gamma_2=\gamma_m=0$. The phase-space trajectory of the oscillator (in figure~\ref{fig:CS}) starts at the origin and moves clock-wise to the resonant position of $b_r/b_0=1$ if $\bar{n}>n_{\rm thr}$. Around here, photons are transferred to the right cavity mode (as seen in figure~\ref{fig:PO}) and the oscillator moves along the other larger circle. Turning back to the resonant position of $b_r/b_0=1$, photons are transferred back to the left mode (as seen in figure~\ref{fig:PO}) and the oscillator moves along the original circle.

%------fig added: damping-dependence transmittivity Tmax---------------
  \begin{figure}[t]
\begin{center}
\includegraphics[height=0.25\textwidth]{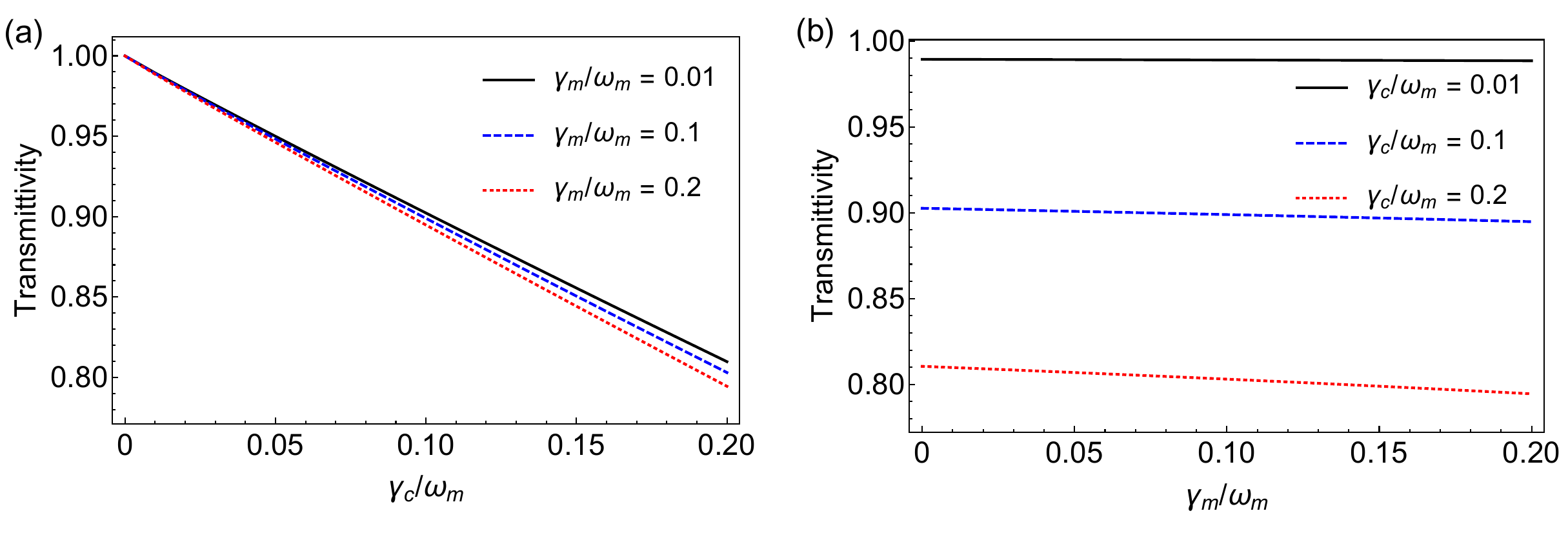}
\end{center}    \caption{Maximum amplitude transmittivity $T_{{\rm max}}$ as a function of mechanical and cavity damping rates, $\gamma_m$ and  $\gamma_1=\gamma_2=\gamma_c$, respectively. (a) $T_{{\rm max}}$ versus $\gamma_{c}/\omega_m$ for fixed $\gamma_{m}/\omega_m=0.01,\,0.1$, and $0.2$. (b) $T_{{\rm max}}$ versus $\gamma_{m}/\omega_m$ for fixed $\gamma_{c}/\omega_m=0.01,\,0.1$, and $0.2$. The parameters are $\bar{n} / n_{{\rm thr}} =5$, $\omega_m/g =  10^{-3}$, and $g/\delta \omega = 10^{-2}$.}
    \label{fig:Trans}
\end{figure}
%----------fig: end-----------------------------------------------------------------------

In the presence of damping, the transfer efficiency decreases, with the dominant
sensitivity set by the cavity damping through the coherence condition
$\bar n(\gamma_c/\omega_m)^2\ll1$ derived below. To quantify this effect, we plot in figure~\ref{fig:Trans} the maximum amplitude transmittivity,
\begin{equation}
T_{{\rm max}}=\mathop{{{\rm max}}}\limits_{t}\frac{|\langle \hat{a}_2(t)\rangle|}{|\langle \hat{a}_1(0)\rangle|},
\end{equation}
as a function of the damping rates of mechanical oscillator $\gamma_m$ and cavities $\gamma_1=\gamma_2=\gamma_c$, using the same parameter settings as in figure~\ref{fig:PO}. Here $T_{\max}$ is obtained by solving the semiclassical Langevin equations~(\ref{QDE11})--(\ref{QDE12}) with
$\gamma_1=\gamma_2=\gamma_c$ and $\gamma_m$ retained. We stress that $T_{\max}$ measures the first-moment amplitude transfer, not the full state
fidelity $F$; figure~\ref{fig:Trans} therefore bounds, rather than equals, the achievable fidelity under
damping.
The results show that the transmittivity is sensitive to the ratio $\gamma_c/\omega_m$, since the dynamic resonant transfer occurs on the mechanical timescale. Thus, in addition to the assumptions discussed above, the resolved-sideband condition $\omega_m\gg \gamma_c$ is required to maintain high transfer efficiency in practice.

The phonon-induced dynamic resonance in a three-mode optomechanical system is at the heart of our approach to QST. Efficient photon transmission between the two cavities is achievable when the key assumptions are satisfied. Specifically, the high-amplitude limit is feasible with a large photon number, as discussed in the subsequent section. The other conditions can be met by appropriately configuring the system parameters. For instance, choosing $g/\delta\omega = 10^{-2}$ for weak optical coupling, $\bar{n}=100$ for high-amplitude regime, and $\nu=0.1$ for adiabaticity, the requirement  $\bar{n}/n_{{\rm thr}}>1$ in Eq.~(\ref{AdbCon}) is satisfied for $\kappa_0/\omega_m \gtrsim 5.6$, corresponding to the ultrastrong optomechanical coupling regime. To also meet the condition for neglecting thermal fluctuations in Eq.~(\ref{negl_thermal}), we require $n_m^{\rm th}\ll 3.13\times 10^5 r^2$, which can be satisfied, for example, for a mechanical frequency in the MHz range at room temperature by choosing $r = 5$, or equivalently, $\kappa_0/\omega_m \approx 28$.

%============Section 4====================================================================
\section{Transfer quantum states with the large number of photons}\label{sec4}
In this section, we propose two schemes to transfer quantum states between two cavities. Initially, the cavities are prepared in a composite state $|\Psi_0\rangle = | \psi \rangle\otimes | 0 \rangle$, where the left cavity is in the state $| \psi \rangle$ and the right cavity is in the vacuum (ground) state $| 0 \rangle$. The local state $| \psi \rangle$ is to be transferred from the left cavity to the right through the phonon-induced dynamic resonance. In the first scheme, we aim that the cavities are in a fixed target state $|\Psi_{\rm fix}\rangle=|0\rangle\otimes|\psi\rangle$, similar to the conventional state-swapping. In the second scheme, we introduce a moving target state $|\Psi_{\rm mov}\rangle=|0\rangle\otimes|\psi_{\rm m}(t)\rangle$, where the transferred state $|\psi_{\rm m}(t)\rangle$ depends  on time in terms of the phase. Under the assumptions required to establish phonon-induced dynamic resonance, we derive expressions for the transfer fidelity (section~\ref{sec4.1}) and  evaluate the efficiency of our schemes by employing Fock states, displaced squeezed states, and Schrödinger cat states (section~\ref{sec4.2}). These states, fundamental to quantum optics, are pivotal for applications in quantum information processing.

%============Subsection 4.1====================================================================
\subsection{Transfer fidelity}\label{sec4.1}
The state transfer is evaluated in terms of fidelity $F$, which is defined by the probability that the cavities initially in $|\Psi_0\rangle$ are found in a target state $|\Psi_{\rm tar}\rangle$ at time $t$, 
%_______Eq: Fid-definition________________________________
\begin{equation}
\label{Fid-def}
F(t)=|\langle \Psi_{\rm tar}|\Psi(t)\rangle|^2,
\end{equation}     
where $|\Psi(t)\rangle=\hat{U}(t)|\Psi_0\rangle$ and the time evolution operator $\hat{U}(t)=\hat{\mathcal{T}}\left[\exp\left(-i\int_0^t \hat{H}_{\rm opt}(\tau)d\tau\right)\right]$ with $\hat{\mathcal{T}}$ being the time-ordering operator and $\hat{H}_{\rm opt}$ being the semiclassical form of $\hat{H}_{\rm eff}$
in which the mechanical mode is considered classical. Here $|\Psi_{\rm tar}\rangle$ is $|\Psi_{\rm fix}\rangle$ for the fixed target state or $|\Psi_{\rm mov}\rangle$ for the moving target state. To distinguish the fidelity between the two schemes, we will also use the subscripts “fix” and “mov” for $F$, corresponding to the fixed and moving target states, respectively.

With the fixed target state $|\Psi_{\rm tar}\rangle:=|\Psi_{\rm fix}\rangle$, by expanding the local state $|\psi\rangle$ as $|\psi\rangle=\sum_nc_n|n\rangle$, where $\{| n \rangle\}$ is a complete set of number states, we get 
%_______Eq: inner product________________________________
\begin{equation}
\label{inn-prod}
\langle\Psi_{\rm tar}|\Psi(t)\rangle=\sum_{n,m=0}^\infty\frac{c^*_nc_m}{\sqrt{n!m!}}\langle 00|\hat{a}^n_2(t)\hat{a}^{\dagger m}_1(0)|00\rangle.
\end{equation}
Here $|00\rangle:=|0\rangle\otimes |0\rangle$. The fidelity $F_{\rm fix}(t)$ in equation~(\ref{Fid-def}) is then obtained by applying the solutions in equation~(\ref{eq:at}) to the operator $\hat{a}_2(t)$ in equation~(\ref{inn-prod}). 

To analytically examine the state-transfer process, we assume in this work that $\gamma_1=\gamma_2=\gamma_m=0$. A brief estimate of the effect of damping rates on the fidelity is provided at the end of this section. A full treatment of quantum dissipation, as well as the effects neglected under the adiabatic approximation and high-amplitude limit, will be presented in future work.

Under this ideal condition, equation~(\ref{eq:at}) gives, as derived in
Appendix~\ref{apdA2},
\begin{equation}
\hat{a}_2(t)=T_{21}(t)\hat{a}_1(0)+T_{22}(t)\hat{a}_2(0), 
\end{equation}
where
%_______Eq: transmittance elements________________________________
\begin{eqnarray}
T_{21}(t)&=&s(t)c(0)e^{-i\xi(t)}-c(t)s(0)e^{i\xi(t)},\label{T21}\\
T_{22}(t)&=&s(t)s(0)e^{-i\xi(t)}+c(t)c(0)e^{i\xi(t)},
\end{eqnarray}
with $\xi(t)=\int_0^t\epsilon(\tau)d\tau$, $c(t)=\sqrt{\left(1+\omega(t)/\epsilon(t)\right)/2}$, $s(t)=\sqrt{\left(1-\omega(t)/\epsilon(t)\right)/2}$, and $\epsilon(t)=\sqrt{\omega^2(t)+g^2}$.  
Thus the fidelity becomes 
%_______Eq: Fidelity_fix________________________________
\begin{equation}\label{Fid}
F_{\rm fix}(t)=\left|\langle\psi|\big( T_{21}(t)\big)^{\hat{n}}|\psi\rangle\right|^2.
\end{equation}
Expressing $T_{21}(t)=|T_{21}(t)|e^{i\theta(t)}$, it is necessary for some $t$ that the transmittivity $|T_{21}|\approx 1$ to achieve high fidelity. From equation~(\ref{T21}), 
%_______Eq: Transmittivity without dissipation |T21|________________________________
\begin{equation}
    |T_{21}(t)|=\left\{s^2(t)c^2(0)+c^2(t)s^2(0)-2s(t)c(t)s(0)c(0)\cos{(2\xi(t))}\right\}^{1/2}.
\end{equation}
For $g/\delta\omega\ll1$ and in the time window $b_r(t)/b_0>1$, where
the population has been transferred to the right cavity, this is approximated by
%_______Eq: approx of |T21|________________________________
\begin{equation}\label{T21-appr}
|T_{21}(t)|\approx \sqrt{1-\left(\frac{g}{\delta\omega}\right)^2\cos^2\xi(t)}.
\end{equation}
In other words, the QST can approach unity fidelity on the condition that the population transfer is nearly complete, $|T_{21}|\approx1$. On the other hand, the fidelity also depends on the phase factor $e^{i\theta(t)\hat n}$, which can oscillate rapidly at a frequency of the order of $g$.

To account for the oscillations in the phase and the fidelity, we introduce an
alternative scheme in which the target state is allowed to acquire the
deterministic phase accumulated during the transfer, called a moving
target state, $|\Psi_{\rm mov}(t)\rangle=|0\rangle\otimes e^{i\theta(t)\hat{n}}|\psi\rangle$. The phase $\theta(t)$, derived from equation~(\ref{T21}), is an analytic function of system
parameters and time, independent of the input state,
%_______Eq: accumulated phase________________________________
\begin{equation}
    \tan\theta(t)=\frac{c(t)s(0)+s(t)c(0)}{c(t)s(0)-s(t)c(0)}\tan\xi(t).
\end{equation}
This corresponds to working in the appropriate rotating frame of the receiver
cavity. If one requires the phase of transferred state to coincide with the initial state, this
deterministic phase can be compensated either by adjusting the detection frame or, equivalently, by applying a calibrated phase
shifter~\cite{PhaseShifter1,PhaseShifter2,add9,add10}. The latter implements a
unitary phase correction;
while practical implementations may introduce small technical noise or
insertion loss, these effects are negligible compared to other noise
sources in typical optomechanical experiments~\cite{add11}.
With this approach, the fidelity depends solely on the magnitude of $T_{21}$, given by
%_______Eq: Fid_moving_____________________________
\begin{equation}
\label{Fid-run}
F_{\rm mov}(t):=|\langle \Psi_{\rm mov}(t)|\Psi(t)\rangle|^2=\left|\langle\psi|\big| T_{21}(t)\big|^{\hat{n}}|\psi\rangle\right|^2.
\end{equation} 
This formulation simplifies the time dependency of the fidelity and effectively addresses the oscillatory behavior.

%============Subsection 4.2====================================================================
\subsection{Examples}\label{sec4.2}
%------------------fig: Fid: Fock state--------------------
\begin{figure}[t]
\begin{center}
\includegraphics[height=0.23\textwidth]{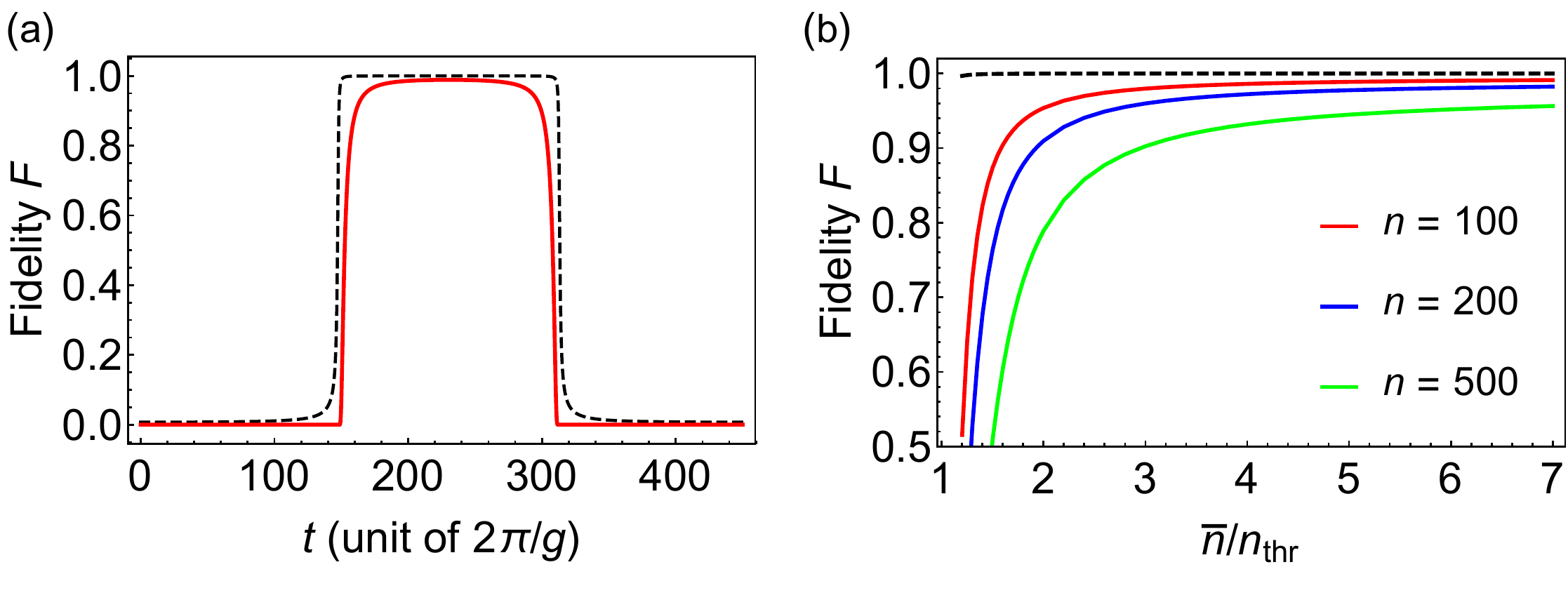}
\end{center}
\caption{
Fidelities of transferring Fock states: (a)  $F_n(t)$ as a function of time   with $n=100$ for $\bar{n}/n_{\rm thr}=5$, and (b) the maximum value of $F_n(t)$ as a function of $\bar{n}/n_{\rm thr}$ for $n=100,200,500$. The parameters $\omega_m/g =  10^{-3}$ and $g/\delta \omega = 10^{-2}$. The dashed black lines represent the transmittivity $|T_{21}(t)|$.}
\label{fig:FidFs}
\end{figure}
%------------------fig: end--------------------

As examples, we first consider the transfer of the Fock state $|\psi\rangle=|n\rangle$. The state must have a large photon number $n\gg1$ to satisfy the high amplitude limit in equation (\ref{HAC_cav}). The fidelity is given by
%_______Eq: Fid_Fock____________________
\begin{equation}\label{Fid-Fock}
F_{{\rm fix},n}(t)=F_{{\rm mov},n}(t)=|T_{21}(t)|^{2n}=:F_n(t).
\end{equation}  
Here the fidelities with the moving target state $F_{\rm mov}(t)$ and the fixed target state $F_{\rm fix}(t)$ are equal to each other. This is because $|\Psi_{\rm mov}(t)\rangle =|0\rangle\otimes e^{i\theta(t)\hat{n}}|n\rangle$ differs from $|\Psi_{\rm fix}\rangle=|0\rangle\otimes|n\rangle$ by a global phase. Using approximation in Eq.~(\ref{T21-appr}), the fidelity in equation~(\ref{Fid-Fock}) becomes
%_______Eq: Fid_Fock: approx____________________
\begin{equation}\label{Fid-Fock-appr}
F_n(t)\approx 1-n\left(\frac{g}{\delta\omega}\right)^2\cos^2\xi(t),
\end{equation}
which approaches unity if $n(g/\delta\omega)^2\ll 1$. This establishes an upper bound on the initial photon number of the local state $|\psi\rangle$, showing
that larger photon numbers necessitate smaller values of $g/\delta\omega$ to
attain high fidelity. To illustrate the possibility of achieving faithful state transfer, we plot the fidelity in equation~(\ref{Fid-Fock}) as a function of time, compared with the transmittivity $|T_{21}|$, in figure~\ref{fig:FidFs}(a). Fidelities of transferring Fock states with different photon numbers are presented in figure~\ref{fig:FidFs}(b). As predicted by equation~(\ref{Fid-Fock-appr}), the fidelity decreases as increasing the initial number of photons $\bar{n}=n$, while keeping the ratio $g/\delta\omega$ constant. Additionally, figure~\ref{fig:FidFs}(b) shows that the fidelity depends on the ratio $\bar{n}/n_{\rm thr}$, however, it rapidly increases to the unity with the ratio. Therefore, this ratio does not need to be very large, noting that an excessively high initial photon number $\bar{n}$ in the left cavity, far exceeding the threshold $n_{\rm thr}$, would violate the adiabatic assumption and destabilize the mechanical oscillator's motion.

\begin{figure}[t]
\begin{center}
\includegraphics[height=0.2\textwidth]{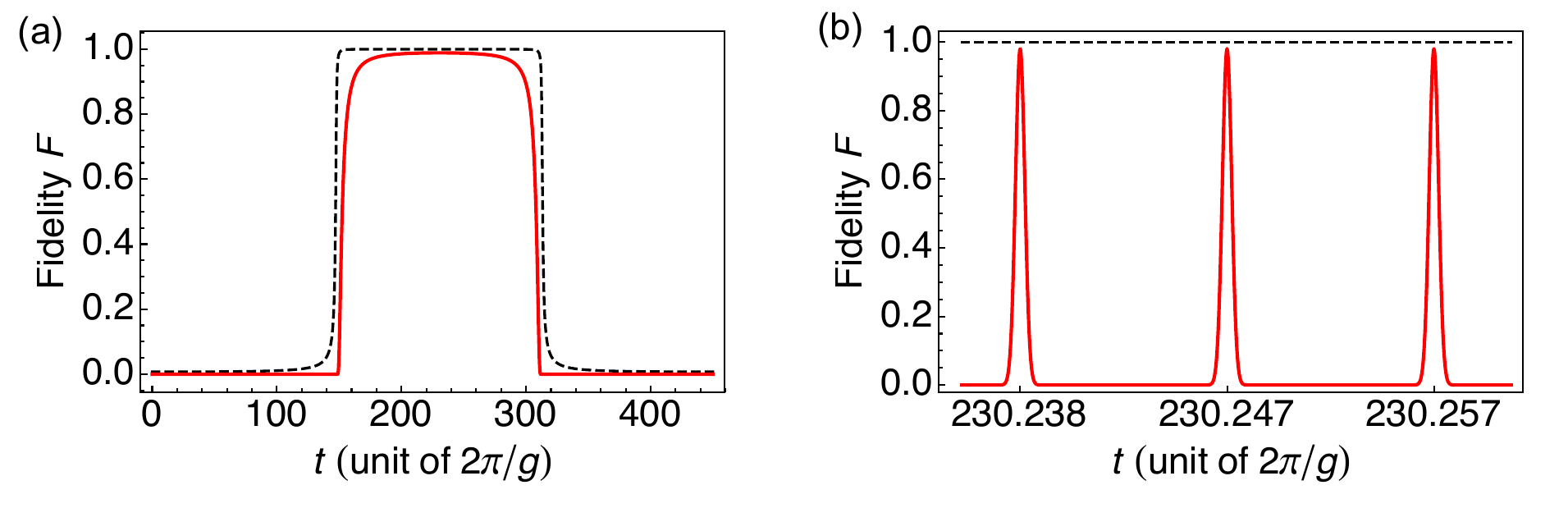}
\end{center}
\caption{
Fidelities of transferring a Schr{\"o}dinger cat state with $|\alpha|=10$: (a) $F_{\rm mov,SC}(t)$ and (b) $F_{\rm fix,SC}(t)$. The parameters $\omega_m/g =  10^{-3}$, $g/\delta \omega = 10^{-2}$, and $\bar{n}/n_{\rm thr}=5$. The dashed black lines represent the transmittivity $|T_{21}(t)|$.}
\label{FFSCS}
\end{figure}

Secondly, we consider Schr{\"o}dinger cat states to transfer~\cite{JLee,SCS1,SCS2,SCS3}
%_______Eq: Sch. cat state: def____________________
\begin{equation}
|\psi\rangle=|\pm\rangle\equiv\mathcal{N}_\pm (|\alpha\rangle\pm|-\alpha\rangle),
\end{equation}
where $|\pm\alpha\rangle$ are coherent states with amplitude $\pm \alpha$ and $\mathcal{N}_\pm=1/\sqrt{2(1\pm\exp(-2|\alpha|^2))}$. The high amplitude limit in equation  (\ref{HAC_cav}) satisfies
%_______Eq: HAL: Sch. cat____________________
\begin{equation}
\label{HAL-SCs}
 1 \pm 2 \abs{\alpha}^2 e^{-2  \abs{\alpha}^2} \ll  \abs{\alpha}
\end{equation}
for the large amplitude $|\alpha|\gg 1$. The fidelities with the fixed target state $|\Psi_{\rm fix}\rangle$ and the moving target state $|\Psi_{\rm mov}(t)\rangle$ are given, respectively, by
%_______Eq: Fid_Sch.cat____________________
\begin{equation}
F_{\rm fix,SC}(t)=\left|\frac{e^{|\alpha|^2T_{21}(t)}\pm e^{-|\alpha|^2T_{21}(t)}}{e^{|\alpha|^2}\pm e^{-|\alpha|^2}}\right|^2
\end{equation}
and
\begin{equation}
F_{\rm mov,SC}(t)=\left(\frac{e^{|\alpha|^2|T_{21}(t)|}\pm e^{-|\alpha|^2|T_{21}(t)|}}{e^{|\alpha|^2}\pm e^{-|\alpha|^2}}\right)^2.
\end{equation}
Using equation~(\ref{T21-appr}), both $F_{\rm fix,SC}(t)$ and $F_{\rm mov,SC}(t)$ approach unity if $|\alpha|^2(g/\delta\omega)^2\ll 1$. Noting that $\bar{n}\approx|\alpha|^2$ for large $|\alpha|$, this condition can be expressed as $\bar{n}(g/\delta\omega)^2\ll 1$, which is the same as that for Fock states. Figure~\ref{FFSCS} displays the fidelities of transferring a Schrödinger cat state with the initial number of photons $\bar{n}\approx 100$. In figure~\ref{FFSCS}(a), the fidelity with the moving target state, $F_{\rm mov,SC}(t)$, consistently approaches unity over a prolonged period (red curve), closely matching the interval when the transmittivity $|T_{21}(t)|$ reaches unity (dashed black curve). On the other hand, in figure~\ref{FFSCS}(b), the fidelity with the fixed target state, $F_{\rm fix,SC}(t)$, reaches unity at specific times $t$ but then decays to zero, even when the population transfer is perfect (indicated by the unity of the dashed black line). The interval between adjacent peaks is approximately $10^{-2}(2\pi/g)$, suggesting the rapid oscillation of $F_{\rm fix,SC}(t)$.
%------------------fig: Fid_Sch.cat--------------------

\begin{figure}[t]
\begin{center}
\includegraphics[height=0.27\textwidth]{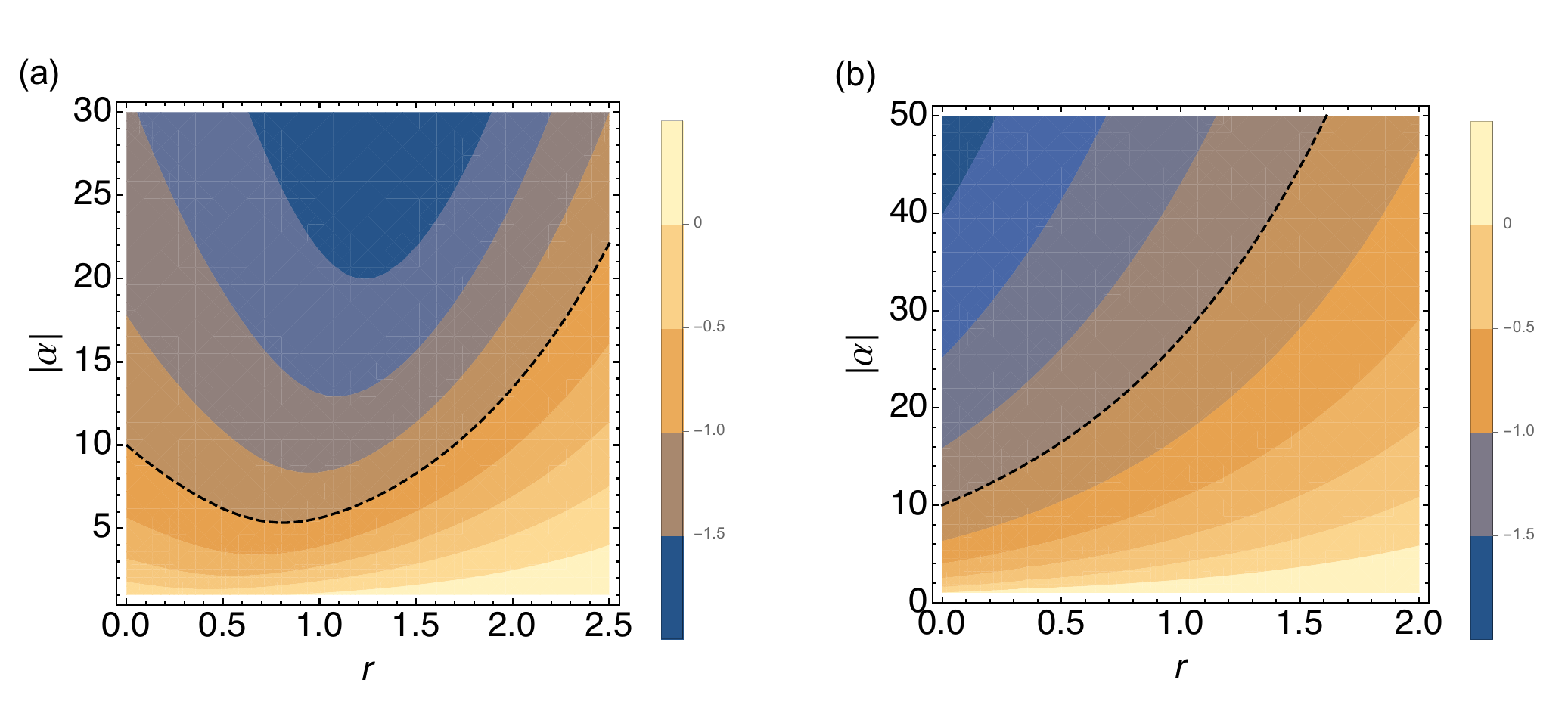}
\end{center}
\caption{Dependence of $\log({\rm HA})$ on the coherent amplitude $|\alpha|$ and squeezing parameter $r$ for (a) $\Delta\varphi=0$ and (b) $\Delta\varphi=\pi$. The dashed contour indicates $\log({\rm HA})=-1$, marking the threshold for the high-amplitude condition of displaced squeezed states, which requires $\log({\rm HA})<-1$.}
\label{fig:HAL}
\end{figure}

\begin{figure}[t]
\begin{center}
\includegraphics[height=0.2\textwidth]{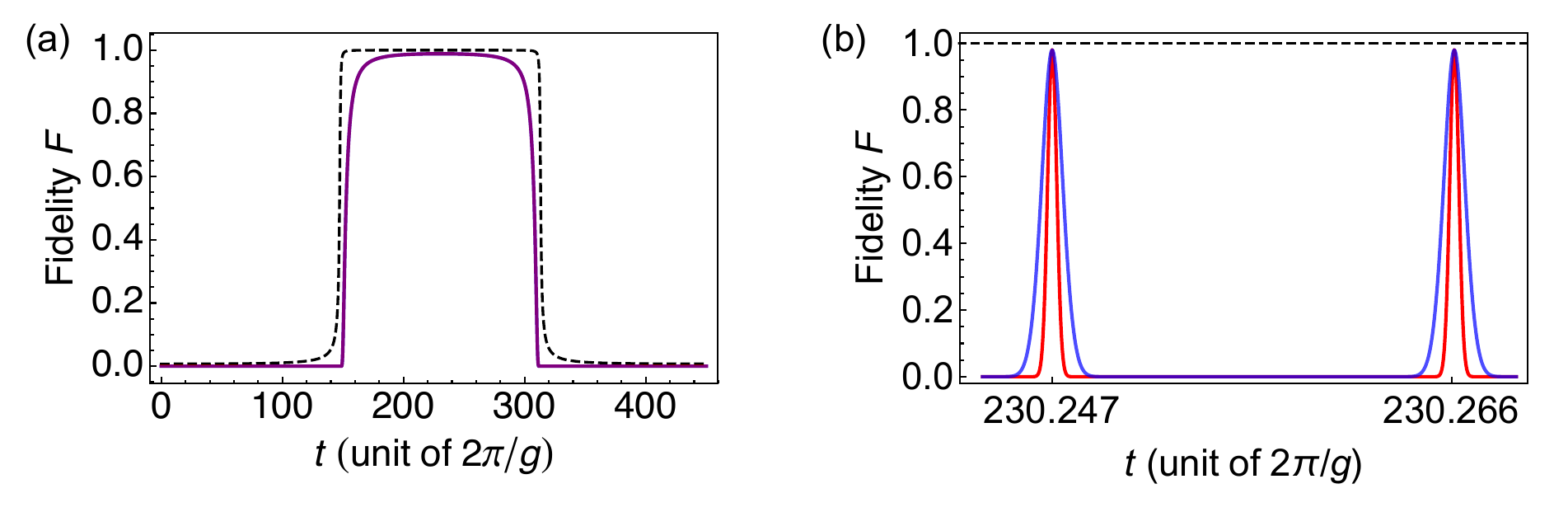}
\end{center}
\caption{
Fidelities of transferring coherent state $|\alpha=10,r=0\rangle$ (red curves) and displaced squeezed state $|\alpha=9.93,r=1\rangle$ (blue curves): (a) $F_{\rm mov,DS}(t)$ and (b) $F_{\rm fix,DS}(t)$. The parameters $\omega_m/g =  10^{-3}$, $g/\delta \omega = 10^{-2}$, and $\bar{n}/n_{\rm thr}=5$. The dashed black lines represent the transmittivity $|T_{21}(t)|$.}
\label{FFCS}
\end{figure}

To evaluate the performance of our model on Gaussian states, we consider the displaced squeezed states~\cite{Knight}
%_______Eq: DS state: def____________________
\begin{equation}\label{CSs}
|\psi\rangle=|\alpha,\eta\rangle\equiv\hat{D}(\alpha)\hat{S}(\eta)|0\rangle,
\end{equation}
where $\hat{D}(\alpha)=\exp(\alpha\hat{a}^\dag-\alpha^*\hat{a})$ is the displacement operator and $\hat{S}(\eta)=\exp\left[\frac{1}{2}\left(\eta^*\hat{a}^2-\eta\hat{a}^{\dag2}\right)\right]$ the squeezing operator with complex amplitudes $\alpha=|\alpha|e^{i\varphi_\alpha}$ and $\eta=re^{i\varphi_\eta}$, respectively. Applying the high amplitude limit equation~(\ref{HAC_cav}) to $|\alpha,\eta\rangle$ results in
%_______Eq: HAL: DS____________________
\begin{equation}\label{HAL-CSs}
{\rm HA}=\frac{\sqrt{|\alpha|^2(\cosh 2r-\sinh 2r \cos\Delta\varphi)+\frac{1}{2}\sinh^2 2r}}{|\alpha|^2+\sinh^2r}\ll 1,
\end{equation}
where $\Delta\varphi=\varphi_\eta-2\varphi_\alpha$. The condition~(\ref{HAL-CSs}) is satisfied by high-amplitude states with large $|\alpha|\gg 1$, along with an appropriate chosen value of $r$, as illustrated in figure~\ref{fig:HAL}. The explicit expressions for fidelities of transferring the displaced squeezed states are provided in  appendix~\ref{apdB}. The numerical results, $F_{\rm mov,DS}(t)$ in equation~(\ref{Fid-DS2}) and $F_{\rm fix,DS}(t)$ in equation~(\ref{Fid-DS1}), for states to transfer $|\alpha=10,r=0\rangle$ and $|\alpha=9.93,r=1\rangle$ are plotted in figure~\ref{FFCS}. Figure~\ref{FFCS}(a) illustrates the achievement of nearly perfect fidelity with the moving target state, while figure~\ref{FFCS}(b) demonstrates the oscillation of fidelity with the fixed target state. \par

%---------fig: HAL: DS----------

%---------fig: Fid_DS----------

We emphasize that all demonstrations consistently exhibit nearly perfect transfer, occurring on the timescale $\tau\sim 1/\omega_m$, for the ideal case without dissipation. When dissipation is included, provided the resolved-sideband $\omega_m\gg \gamma_c$ holds, the coherence condition $\tau \gamma_c \ll 1$ keeps the amplitude loss small. The fidelity requirement depends on the input state. For instance, the transfer of a coherent state remains essentially lossless as long as $\bar{n}(\tau\gamma_c)^2\sim \bar{n}(\gamma_c/\omega_m)^2 \ll 1$~\cite{CMCTheo3}. A stricter condition may be required for non-Gaussian states, e.g.,
$\bar n(\gamma_c/\omega_m)\ll1$; while more demanding, such a condition
can still be satisfied in principle. This behavior stands in clear contrast to linearized approaches, where high fidelity is limited essentially to single-photon states due to the inherent trade-off constraint  $\tau \gamma_c \sim 1$~\cite{CMExp1,SCExp6,CMExp3,CMCTheo1,CMCTheo3,CMCTheo4}.\par

This section demonstrates how phonon-induced dynamic resonance enables efficient quantum state transfer. Once population transfer is achieved, the phase coherence is preserved up to a deterministic, state-independent phase evolution, which can be compensated by the phase correction discussed above. Transfer fidelities up to 0.99 are attainable for the system parameters used throughout this work, requiring operation in the ultrastrong optomechanical coupling regime. Several modern architectures, such as atom-based and nanowire-in-fiber-cavity optomechanical platforms, have approached the single-photon (ultra)strong optomechanical coupling regime~\cite{Murch,Burgwal,add6}. Extending this progress to simultaneously satisfy the resolved-sideband condition required in the presence of dissipation remains an active direction of current research~\cite{add14}. Superconducting circuit optomechanical platforms and circuit-QED-based optomechanical-like platforms provide particularly promising routes toward this regime~\cite{Lemonde,add12,add13,Manninen,add16}.

Beyond this near-ideal benchmark, the assumptions can be relaxed toward more conventional regimes. Since weak hopping, the most essential assumption, is readily satisfied in practice and the high-amplitude condition pertains only to the quantum state, adiabaticity can be relaxed with a slight loss in fidelity. Alternatively, without relaxing any assumption, recent developments in counter-diabatic control~\cite{add7,add15,add8} suggest that the adiabatic requirement can be lifted and, consequently, the requirement of strong optomechanical coupling can also be relaxed, opening the way to applications in more conventional optomechanical settings.

%============Section 5====================================================================
\section{Conclusion}\label{con}\label{sec5}

We propose a novel method to transfer multi-photon quantum states between two optical cavities mediated by a mechanical oscillator in a three-mode optomechanical system. The system operates in the nonlinear optomechanical regime under three key assumptions: the high-amplitude limit of the mechanical oscillator, weak coupling between the optical cavities, and the adiabatic approximation. These conditions establish phonon-induced dynamic resonance, which underlies the quantum-state transfer. A defining strength of our scheme is its capacity to transfer high-amplitude quantum states, a capability unattainable in approaches relying on linearized optomechanical interactions. Efficiently transferring such high-amplitude quantum states is essential for advancing quantum computing with continuous variables~\cite{Braunstein, Fukui}.

We examine two transfer schemes with fixed and moving target states. For the fixed target, the fidelity rapidly oscillates because of the deterministic phase accumulated during the transfer. In the moving-target scheme, this phase is compensated either by choosing the appropriate rotating frame in detection or by applying a calibrated optical phase shifter. The phase correction is state independent and unitary, allowing high fidelity to be maintained over a timescale comparable to that of the mechanical oscillator.

In this work, the analytical fidelity treatment neglects quantum and thermal fluctuations of the mechanical oscillator, as well as damping and noise. Under the high-amplitude condition, 
the leading dynamics are governed by the coherent classical oscillation, while quantum corrections associated with photon blockade, nonlinear photon correlations, cavity--mechanical entanglement, and multi-sideband resonances~\cite{MTK} are expected to be perturbative. A full quantum treatment including these effects and dissipation is left for future work.

%\section*{References}

%
% Each of the commands below will create an unnumbered section with the appropriate heading.
% Remove any sections that are not relevant for your article.
% All sections except suppdata will be removed if the [anonymous] option is used.
% See iopjournal-guidelines.pdf for more information.
%

\section*{Acknowledgment}
This work was supported by the National Research Foundation of Korea (NRF) grant funded by the Korea government (Ministry of Science and ICT) (Grant No.\ 2023M3K5A\allowbreak1094813) and by Creation of the Quantum Information Science R\&D Ecosystem (Grant No. RS-2023-NR068116) through the National Research Foundation of Korea (NRF) funded by the Korean government (Ministry of Science and ICT).

\appendix 
%============Apd A====================================================================
\section*{Appendix}
\section{Adiabatic approximation} \label{apdA}

We derive the adiabatic approximation to the cavity dynamics. We perform an eigendecomposition for the cavity modes to obtain ${\bf \Lambda}(t) = {\bf W}^{-1}(t) {\bf M}_a(t) {\bf W}(t) $ for some invertible matrix $ {\bf W}(t) $ so that ${\bf \Lambda}(t) $ is a diagonal matrix at every instance time $t$; the column vectors of $ {\bf W}(t) $ are instantaneous eigenvectors of the dynamic matrix ${\bf M}_a(t)$ in equation~(\ref{DM}), 
\begin{equation}\label{EE}
{\bf M}_a(t)  {\bf w}_{j}(t) = \lambda_{j}(t) {\bf w}_{j}(t),
\end{equation}
where ${\bf w}_{j}(t)$ is the $j$-th column vector of $ {\bf W}(t) $ and $\lambda_{j}(t)$ is the $j$-th diagonal element of ${\bf \Lambda}(t)$. 
%Note $c^{2}(t) + s^{2}(t) = 1$ and $\bold{W}^{-1}(t) = \bold{W}^{T}(t) $ with the orthonormality $\bold{W}^{T}(t)  \bold{W}(t) = \mathbb{I}$. 

Applying ${\bf W}^{T}(t)$ from the left, equation~(\ref{QDE11}) is transformed to the dynamic equation for the eigenmode vector
\begin{equation}\label{QDE2}
 i \dot{{\bf a}}_\pm(t) = {\bf \Lambda}(t)  {\bf a}_\pm+i{\bf f}^{{\rm in}}_\pm(t),
\end{equation}
where ${\bf a}_\pm(t)={\bf W}^T(t){\bf a}(t)$, ${\bf f}^{\rm in}_\pm(t)={\bf W}^T(t){\bf f}^{\rm in}_a(t)$. 
Here, we use $ {\bf w}^{T}_{j}(t) \dot{{\bf w}}_{j}(t) = 0$ for all $j$. We neglect $ {\bf w}^{T}_{j}(t) \dot{{\bf w}}_{k}(t)$ by the adiabatic approximation \cite{Adia2} 
\begin{equation}\label{ADC1}
\nu = {\rm Max}_{t} \abs{  \frac{{\bf w}^{T}_{j}(t) \dot{{\bf w}}_{k}(t)}{\lambda_{j}(t) - \lambda_{k}(t)}} \ll 1
\end{equation}
for  $j \neq k$. Using the expressions for quantities associated with $\nu$, given in sections \ref{sec2} and \ref{sec3}, the adiabatic is rephrased to be 
\begin{equation}\label{ADC2}
\nu  \approx \frac{1}{8}\frac{\bar{n}}{n_{\rm thr}}\frac{\omega_m \delta \omega}{g^{2}} \ll 1.
\end{equation}

Accordingly, equation~(\ref{QDE11}) is transformed to the dynamic equation for eigen modes of annihilation operators ${\bf a}_\pm(t)=\big(\hat{a}_+(t),\hat{a}_-(t)\big)^T$,
%_______Eq: Langevin eqs for cavities in adiabatic approx________________________________
\begin{equation}\label{eig}
i\dot{{\bf a}}_\pm(t)={\bf \Lambda}(t){\bf a}_\pm(t)+i{\bf f}^{{\rm in}}_\pm(t).
\end{equation}
For $\gamma_1=\gamma_2$, the eigenvalues are given by $\lambda_\pm(t)=\pm\epsilon(t)-i(\gamma_1+\gamma_2)/2$, where $\epsilon(t)=\sqrt{\omega^2(t)+g^2}$. The transformation ${\bf W}(t)$ is expressed as
%_______Eq: transformation matrix W________________________________
\begin{equation}\label{Wt}
{\bf W}(t)=\left(
\begin{array}{cc}
c(t) & -s(t) \\
s(t) & c(t)
\end{array}
\right)
\end{equation}
with $c(t)=\sqrt{\left(1+\omega(t)/\epsilon(t)\right)/2}$ and $s(t)=\sqrt{\left(1-\omega(t)/\epsilon(t)\right)/2}$. The solution to equation~(\ref{eig}) is given by
%_______Eq: solution a(t)________________________________
\begin{equation}
\label{eq:at}
{\bf a}_\pm(t)=e^{-i{\bf\Xi}(t,0)}{\bf a}_\pm(0)+\int^t_0 d\tau e^{-i{\bf\Xi}(t,\tau)}{\bf f}^{\rm in}_\pm(\tau),
\end{equation}
where ${\bf\Xi}(t,\tau)=\int^t_\tau dt' {\bf\Lambda}(t')$. It is remarkable that the two coupled cavities are described by two independent eigen modes. On the other hand, each eigen mode remains coupled with the oscillator in a way that its frequency depends temporally on the amplitude of oscillator, which is the solution to equation~(\ref{QDE12}),
%_______Eq: solution b(t)________________________________
\begin{eqnarray}\label{bt}
b(t)&=& e^{-i\xi_m(t,0)}b(0)+i\int^t_0d\tau e^{-i\xi_m(t,\tau)}\kappa_0\langle\Delta\hat{n}(\tau)\rangle\nonumber\\
&&\approx e^{-i\xi_m(t,0)}b(0)\nonumber\\
&&\quad+i\int^t_0d\tau e^{-i\xi_m(t,\tau)}\kappa_0(\omega(\tau)/\epsilon(\tau))\langle\Delta\hat{n}_\pm(\tau)\rangle,
\end{eqnarray}
where $\xi_m(t,\tau)=(\omega_m-i\gamma_m)(t-\tau)$, $\Delta\hat{n}_\pm(\tau)=\hat{n}_+(\tau)-\hat{n}_-(\tau)$ with $\hat{n}_\pm=\hat{a}_\pm^\dag\hat{a}_\pm$. Here we neglected the terms including $\hat{a}_\pm^\dag(\tau)\hat{a}_\mp(\tau)$, by the rotating wave approximation, as they are fast rotating with 
$|\epsilon(\tau)|\gg|\omega_m-i\gamma_m|\simeq\omega_m$.\par

\section{Ideal transmittivity in adiabatic approximation} \label{apdA2}
By assuming no noise in equation~(\ref{eq:at}), we obtain
%_______Eq: a(t) without dissipation________________________________
\begin{equation}\label{vat}
\left(
\begin{array}{c}
\hat{a}_1(t)\\
\hat{a}_2(t)
\end{array}
\right)={\bf T}(t)
\left(
\begin{array}{c}
\hat{a}_1(0)\\
\hat{a}_2(0)
\end{array}
\right),
\end{equation}
where ${\bf T}(t)={\bf W}(t)e^{-i{\bf \Xi}(t,0)}{\bf W}^T(0)$. The transmittance matrix ${\bf T}(t)$ is composed of elements
%_______Eq: transmittance elements________________________________
\begin{eqnarray}
T_{22}(t)&=&T^*_{11}(t)=s(t)s(0)e^{-i\xi(t)}+c(t)c(0)e^{i\xi(t)},\\
T_{21}(t)&=&-T^*_{12}(t)=s(t)c(0)e^{-i\xi(t)}-c(t)s(0)e^{i\xi(t)},\label{B3}
\end{eqnarray}
where $s(t)$ and $c(t)$ are defined by equation~(\ref{Wt}), and $\xi(t)=\int_0^t\epsilon(\tau)d\tau$. In particular, 
\begin{equation}
\hat{a}_2(t)=T_{21}(t)\hat{a}_1(0)+T_{22}(t)\hat{a}_2(0),
\end{equation} 
so that $|T_{21}(t)|$ characterizes the transmittivity from cavity 1 to cavity 2.

From equation~(\ref{B3}), we have
%_______Eq: Transmittivity without dissipation |T21|________________________________
\begin{equation}
    |T_{21}(t)|=\left\{s^2(t)c^2(0)+c^2(t)s^2(0)-2s(t)c(t)s(0)c(0)\cos{(2\xi(t))}\right\}^{1/2}.
\end{equation}
For $g/\delta\omega\ll 1$, this can be approximated as
%_______Eq: approx of |T21|________________________________
\begin{equation}
    |T_{21}(t)| \approx \left\{s^2(t)-\frac{g}{\delta \omega}s(t)c(t)\cos{2\xi(t)}+\frac{1}{4}\left(\frac{g}{\delta\omega}\right)^2\left(c^2(t)-s^2(t)\right)\right\}^{1/2}.\label{B6}
\end{equation}
Given that $g/\delta\omega\ll 1$ and $0<s(t),c(t)<1$, $|T_{21}(t)|\approx 1$ holds if
\begin{equation}
    s^2(t)=\frac{1}{2}\left(1-\frac{{\rm sign}[\omega(t)]}{\sqrt{1+g^2/\omega^2(t)}}\right)\approx 1,
\end{equation}
which requires $g^2/\omega^2(t)\ll 1$ and $\omega(t)<0$. Since $\omega(t)=\delta\omega(1-b_r(t)/b_0)$, the condition $g^2/\omega^2(t)\ll 1$ is readily satisfied under the assumption $g/\delta\omega\ll 1$, provided $b_r(t)\neq b_0$. Meanwhile the requirement $\omega(t)<0$ is equivalent to $b_r(t)>b_0$, consistent with the scenario where population transfer is complete, as illustrated in figures~\ref{fig:CS} and \ref{fig:PO}. 

In this time window, the transmittivity can be further estimated as
\begin{equation}\label{B8}
|T_{21}(t)|
\approx
\sqrt{1-\varepsilon^2\cos^2\xi(t)}.
\end{equation}
Since $\xi(t)$ varies rapidly compared with the mechanical motion, averaging
over this fast oscillation and expanding to the order of $\varepsilon^2$
gives
\begin{equation}\label{T21-ave}
\overline{|T_{21}|}
\approx
1-\frac{\varepsilon^2}{2}\overline{\cos^2\xi(t)}
=
1-\frac{\varepsilon^2}{4},
\end{equation}
where $\overline{\cos^2\xi(t)}=1/2$. This result provides the estimate of
the ideal transmission used in equation~(\ref{Tave}).

%============Apd B====================================================================
\section{Fidelity of transferring the displaced squeezed state} \label{apdB}

Substituting $|\psi\rangle=|\alpha,\eta\rangle$ into equations~(\ref{Fid}) and~(\ref{Fid-run}), we obtain
\begin{eqnarray}
F_{\rm fix,DS}(t)&=&\mathcal{C}(t)\big|\langle \alpha,\eta|\alpha'(t),\eta'(t)\rangle\big|^2,\label{Fid-DS1}\\
F_{\rm mov,DS}(t)&=&\mathcal{C}(t)\big|\langle \alpha e^{i\theta(t)},\eta e^{i2\theta(t)}|\alpha'(t),\eta'(t)\rangle\big|^2,\label{Fid-DS2}
\end{eqnarray}
where
\begin{eqnarray}
\eta'(t)&=&r'(t)e^{i\varphi_\eta'(t)},\\
r'(t)&=&{\rm arctanh}[|T_{21}(t)|^2\tanh r],\\
\varphi_\eta'&=&\varphi_\eta+2\theta(t),\\
\alpha'(t)&=&T_{21}(t)\alpha+\beta'(t),\\
\beta'(t)&=&\beta\cosh r'(t)-\beta^*e^{i\varphi_\eta'(t)}\sinh r'(t),\\
\beta(t)&=&e^{i\varphi_\eta}\alpha^*T_{21}(t)(1-|T_{21}(t)|^2)\tanh r\cosh r'(t).
\end{eqnarray}
Here, the coefficient
\begin{eqnarray}
    \mathcal{C}(t)&=&\frac{\cosh r'(t)}{\cosh r}\exp\left[-|\alpha|^2\left(1-|T_{21}(t)|^2\right)\right]\nonumber\\
    &&\quad\times\exp\left[-|\alpha|^{2}(1-|T_{21}(t)|^2)^2\tanh r\cos\Delta\varphi\right]\nonumber\\
    &&\quad\times\exp\left[-|\alpha|^{2}|T_{21}(t)|^2(1-|T_{21}(t)|^2)^2\right.\nonumber\\
    &&\qquad\qquad \times \left.\tanh^2 r\cosh r'(t)\sinh r'(t)\cos\Delta\varphi\right]\nonumber\\
    &&\quad\times\exp\left[|\alpha|^2|T_{21}(t)|^2(1-|T_{21}(t)|^2)^2\tanh^2r\cosh^2r'\right]
\end{eqnarray}
and the inner-product between displaced squeezed states ~\cite{CSstate}
\begin{equation}
\langle \alpha_1,\eta_1|\alpha_2,\eta_2\rangle=\frac{1}{\sqrt{\sigma}}\exp\left[\frac{\eta_{21}\eta_{12}^*}{2\sigma}+\frac{1}{2}(\alpha_2\alpha_1^*-\alpha_2^*\alpha_1)\right],
\end{equation}
where
\begin{eqnarray}
\eta_{j}&=&r_j e^{i\varphi_{\eta_j}},\\
\sigma&=&\cosh r_2 \cosh r_1-e^{i(\varphi_{\eta_2}-\varphi_{\eta_1})}\sinh r_2\sinh r_1,\\
\eta_{ij}&=&(\alpha_i-\alpha_j)\cosh r_i+  (\alpha_i^*-\alpha_j^*)e^{i\varphi_{\eta_i}}\sinh r_i.
\end{eqnarray}

For the special case of coherent state ($\eta=0$), equations.~(\ref{Fid-DS1}) and~(\ref{Fid-DS2}) reduce to
\begin{eqnarray}
F_{\rm fix,C}(t)&=&\exp\left\{-2|\alpha|^2\left[1-|T_{21}(t)|\cos\theta(t)\right]\right\},\\
F_{\rm mov,C}(t)&=&\exp\left[-2|\alpha|^2\left(1-|T_{21}(t)|\right)\right].
\end{eqnarray}

%\funding{Sample text inserted for demonstration.}
% This section is a list of funder names and grant numbers

%\roles{Sample text inserted for demonstration.}
% List author names and the contributions made to the article, using terms from the NISO Contributor Roles Taxonomy (CRediT) https://credit.niso.org

%\data{Sample text inserted for demonstration.}
% For more information on IOP Publishing's research data policy see: https://publishingsupport.iopscience.iop.org/questions/research-data/

%\suppdata{Sample text inserted for demonstration.}

%\section*{References}
\bibliographystyle{quantum}
\bibliography{ref}

@PREAMBLE{
 "\providecommand{\noopsort}[1]{}" 
 # "\providecommand{\singleletter}[1]{#1}%" 
}

@ARTICLE{QIP1,
   author       = {DiVincenzo, D. P.},
   title        = {The physical implementation of quantum computation},
   journal      = {Fortschr. Phys.},
   volume       = {48}, 
   pages        = {771--783},
   year         = {2000},
   doi = {10.1002/1521-3978(200009)48:9/11<771::AID-PROP771>3.0.CO;2-E},
}

@ARTICLE{QIP2,
   author       = "Gisin, N.  and Thew, R.", 
   title        = "Quantum communication", 
   journal      = "Nat. Photon.", 
   volume       = "1", 
   pages        = "165--171", 
   year         = "2007", 
   doi = "10.1038/nphoton.2007.22",
}

@ARTICLE{QIP3,
   author       = "Kimble, H. J.", 
   title        = "The quantum internet", 
   journal      = "Nature (London)", 
   volume       = "453", 
   pages        = "1023--1030", 
   year         = "2008", 
   doi = "10.1038/nature07127",
}

@ARTICLE{QIP4,
   author       = "Awschalom, D. and others", 
   title        = "Development of quantum interconnects ({Q}u{IC}s) for next-generation information technologies", 
   journal      = "PRX Quantum", 
   volume       = "2", 
   pages        = "017002", 
   year         = "2021", 
   doi = "10.1103/PRXQuantum.2.017002"
}

@ARTICLE{QIP5,
   author       = "Moody, G. and others", 
   title        = "2022 Roadmap on integrated quantum photonics", 
   journal      = "J. Phys. Photonics", 
   volume       = "4", 
   pages        = "012501", 
   year         = "2022", 
   doi = "10.1088/2515-7647/ac1ef4",
}

@ARTICLE{QT1,
   author       = "Bennett, C. H. and Brassard, G. and Cr{\'{e}}peau, C. and Jozsa, R. and Peres, A. and Wootters, W. K.", 
   title        = "Teleporting an unknown quantum state via dual classical and {E}instein-{P}odolsky-{R}osen channels", 
   journal      = "Phys. Rev. Lett.", 
   volume       = "70", 
   pages        = "1895", 
   year         = "1993", 
   doi = "10.1103/PhysRevLett.70.1895",
}

@ARTICLE{QT2,
   author       = "Braunstein, S. L. and Kimble, H. J.", 
   title        = "Teleportation of continuous quantum variables", 
   journal      = "Phys. Rev. Lett.", 
   volume       = "80", 
   pages        = "869", 
   year         = "1998", 
   doi = "10.1103/PhysRevLett.80.869",
}

@ARTICLE{QT3,
   author       = "Lee, J. and Kim, M. S. and Park, Y. J. and Lee, S.", 
   title        = "Partial teleportation of entanglement in a noisy environment", 
   journal      = "J. Mod. Opt.", 
   volume       = "47", 
   pages        = "2151--2164", 
   year         = "2000", 
   doi ="10.1080/09500340008235138",
}

@ARTICLE{QT4,
   author       = "Lee, J. and Kim, M. S.", 
   title        = "Entanglement teleportation via {W}erner states", 
   journal      = "Phys. Rev. Lett.", 
   volume       = "84", 
   pages        = "4236", 
   year         = "2000", 
   doi="10.1103/PhysRevLett.84.4236",
}

@ARTICLE{QT5,
   author       = "Lee, J. and Min, H. and Oh, S. D.", 
   title        = "Multipartite entanglement for entanglement teleportation", 
   journal      = "Phys. Rev. A", 
   volume       = "66", 
   pages        = "052318", 
   year         = "2002", 
   doi = "10.1103/PhysRevA.66.052318",
}

@ARTICLE{Swap,
   author       = "Cirac, J. I. and Zoller, P. and Kimble, H. J. and Mabuchi, H.", 
   title        = "Quantum state transfer and entanglement distribution among distant nodes in a quantum network", 
   journal      = "Phys. Rev. Lett.", 
   volume       = "78", 
   pages        = "3221", 
   year         = "1997", 
   doi = "10.1103/PhysRevLett.78.3221",
}

@ARTICLE{APP,
   author       = "Pellizzari, T.", 
   title        = "Quantum networking with optical fibres", 
   journal      = "Phys. Rev. Lett.", 
   volume       = "79", 
   pages        = "5242", 
   year         = "1997", 
   doi = "10.1103/PhysRevLett.79.5242",
}

@ARTICLE{STIRAP1,
   author       = "Gaubatz, U. and Rudecki, P. and Schiemann, S. and Bergmann, K.", 
   title        = "Population transfer between molecular vibrational levels by stimulated {R}aman scattering with partially overlapping laser fields. {A} new concept and experimental results", 
   journal      = "J. Chem. Phys.", 
   volume       = "92", 
   pages        = "5363--5376", 
   year         = "1990", 
   doi = "10.1063/1.458514",
}

@ARTICLE{STIRAP2,
   author       = "Bergmann, K. and Theuer,  H. and Shore,  B. W.", 
   title        = "Coherent population transfer among quantum states of atoms and molecules", 
   journal      = "Rev. Mod. Phys.", 
   volume       = "70", 
   pages        = "1003", 
   year         = "1998", 
   doi = "10.1103/RevModPhys.70.1003",
}

@ARTICLE{STIRAP3,
   author       = "Vitanov, N. V. and Rangelov, A. A. and Shore, B. W. and Bergmann, K.", 
   title        = "Stimulated {R}aman adiabatic passage in physics, chemistry, and beyond", 
   journal      = "Rev. Mod. Phys.", 
   volume       = "89", 
   pages        = "015006", 
   year         = "2017", 
   doi = "10.1103/RevModPhys.89.015006",
}

@ARTICLE{atom1a,
   author       = "Lukin, M. D. and Yelin, S. F. and Fleischhauer, M.", 
   title        = "Entanglement of atomic ensembles by trapping correlated photon states", 
   journal      = "Phys. Rev. Lett.", 
   volume       = "84", 
   pages        = "4232", 
   year         = "2000", 
   doi = "10.1103/PhysRevLett.84.4232",
}

@ARTICLE{atom1b,
   author       = "Fleischhauer, M. and Lukin, M. D.", 
   title        = "Dark-state polaritons in electromagnetically induced transparency", 
   journal      = "Phys. Rev. Lett.", 
   volume       = "84", 
   pages        = "5094", 
   year         = "2000", 
   doi = "10.1103/PhysRevLett.84.5094",
}

@ARTICLE{atom1c,
   author       = "Phillips, D. F. and Fleischhauer, A. and Mair, A. and Walsworth, R. L. and Lukin, M. D.", 
   title        = "Storage of Light in Atomic Vapor", 
   journal      = "Phys. Rev. Lett.", 
   volume       = "86", 
   pages        = "783", 
   year         = "2001", 
   doi = "10.1103/PhysRevLett.86.783",
}

@ARTICLE{atom1d,
   author       = "Mair, A. and Hager, J. and Phillips, D. F. and Walsworth, R. L. and Lukin, M. D.", 
   title        = "Phase coherence and control of stored photonic information", 
   journal      = "Phys. Rev. A", 
   volume       = "65", 
   pages        = "031802(R)", 
   year         = "2002", 
   doi = "10.1103/PhysRevA.65.031802",
}

@ARTICLE{atom2a,
   author       = "Lloyd, S. and Shahriar, M. S. and Shapiro, J. H. and Hemmer, P. R.", 
   title        = "Long Distance, Unconditional Teleportation of Atomic States via Complete {B}ell State Measurements", 
   journal      = "Phys. Rev. Lett.", 
   volume       = "87", 
   pages        = "167903", 
   year         = "2001", 
   doi = "10.1103/PhysRevLett.87.167903",
}

@ARTICLE{atom2b,
   author       = "Leent, T. V. and Bock,  M. and Fertig, F. and Garthoff, R. and Eppelt, S. and Zhou, Y. and Malik, P. and Seubert, M. and Bauer, T. and Rosenfeld, W. and Zhang, W. and Becher, C. and Weinfurter, H.", 
   title        = "Entangling single atoms over 33 km telecom fibre", 
   journal      = "Nature", 
   volume       = "607", 
   pages        = "69--73", 
   year         = "2022", 
   doi = "10.1038/s41586-022-04764-4",
}

@ARTICLE{atom3a,
   author       = "Julsgaard, B. and Kozhekin, A. and Polzik, E. S.", 
   title        = "Experimental long-lived entanglement of two macroscopic objects", 
   journal      = "Nature", 
   volume       = "413", 
   pages        = "400--403", 
   year         = "2001", 
   doi = "10.1038/35096524",
}

@ARTICLE{atom3b,
   author       = "Duan, L.-M. and Lukin, M. D. and Cirac, J. I. and Zoller, P.", 
   title        = "Long-distance quantum communication with atomic ensembles and linear optics", 
   journal      = "Nature", 
   volume       = "414", 
   pages        = "413--418", 
   year         = "2001", 
   doi = "10.1038/35106500",
}

@ARTICLE{atom3c,
   author       = "Razavi, M. and Shapiro, J. H.", 
   title        = "Long-distance quantum communication with neutral atoms", 
   journal      = "Phys. Rev. A", 
   volume       = "73", 
   pages        = "042303", 
   year         = "2006", 
   doi = "10.1103/PhysRevA.73.042303",
}

@ARTICLE{atom4,
   author       = "Sherson, J. F. and Krauter, H. and Olsson, R. K. and Julsgaard, B. and Hammerer, K. and Cirac, I. and Polzik, E. S.", 
   title        = "Quantum teleportation between light and matter", 
   journal      = "Nature", 
   volume       = "443", 
   pages        = "557--560", 
   year         = "2006", 
   doi = "10.1038/nature05136",
}

@ARTICLE{atom5a,
   author       = "Yao, W. and Liu, R. B. and Sham, L. J.", 
   title        = "Theory of control of the spin-photon interface for quantum networks", 
   journal      = "Phys. Rev. Lett.", 
   volume       = "95", 
   pages        = "030504", 
   year         = "2005", 
   doi = "10.1103/PhysRevLett.95.030504",
}

@ARTICLE{atom5b,
   author       = "Ritter, S. and N{\"{o}}lleke,  C. and Hahn, C. and Reiserer, A. and Neuzner, A. and Uphoff, M. and M{\"{u}}cke, M. and Figueroa, E. and Bochmann, J. and Rempe, G.", 
   title        = "An elementary quantum network of single atoms in optical cavities", 
   journal      = "Nature", 
   volume       = "484", 
   pages        = "195--200", 
   year         = "2012", 
   doi = "10.1038/nature11023",
}

@ARTICLE{atom5c,
   author       = "Pe{\~{n}}as, G. F. and Puebla, R. and Garc{\'{i}}a-Ripoll, J. J.", 
   title        = "Improving quantum state transfer: correcting non-{M}arkovian and distortion effects", 
   journal      = "Quantum Sci. Technol.", 
   volume       = "8", 
   pages        = "045026", 
   year         = "2023", 
   doi = "10.1088/2058-9565/acf60a",
}

@ARTICLE{atom6a,
   author       = "Serafini, A. and Mancini, S. and Bose, S.", 
   title        = "Distributed Quantum Computation via Optical Fibers", 
   journal      = "Phys. Rev. Lett.", 
   volume       = "96", 
   pages        = "010503", 
   year         = "2006", 
   doi = "10.1103/PhysRevLett.96.010503",
}

@ARTICLE{atom6b,
   author       = "Kato, S. and N\'{e}met, N. and Senga, K. and Mizukami, S. and Huang, X. and Parkins, S. and Aoki, T.", 
   title        = "Observation of dressed states of distant atoms with delocalized photons in coupled-cavities quantum electrodynamics", 
   journal      = "Nat. Commum.", 
   volume       = "10", 
   pages        = "1160", 
   year         = "2019", 
   doi = "10.1038/s41467-019-08975-8",
}

@ARTICLE{atom7a,
   author       = "Vogell, B. and Vermersch, B. and Northup, T. E. and Lanyon, B. P. and Muschik, C. A.", 
   title        = "Deterministic quantum state transfer between remote qubits in cavities", 
   journal      = "Quantum Sci. Technol.", 
   volume       = "2", 
   pages        = "045003", 
   year         = "2017", 
   doi = "10.1088/2058-9565/aa868b",
}

@ARTICLE{atom7b,
   author       = "Gogyan, A. and Malakyan, Yu.", 
   title        = "Deterministic quantum state transfer between remote atoms with photon-number superposition states", 
   journal      = "Phys. Rev. A", 
   volume       = "98", 
   pages        = "052304", 
   year         = "2018", 
   doi = "10.1103/PhysRevA.98.052304",
}

@ARTICLE{spin1a,
   author       = "Bose, S.", 
   title        = "Quantum communication through an unmodulated spin chain", 
   journal      = "Phys. Rev. Lett.", 
   volume       = "91", 
   pages        = "207901", 
   year         = "2003", 
   doi = "10.1103/PhysRevLett.91.207901"
}

@ARTICLE{spin1b,
   author       = "Lyakhov, A. and Bruder, C.", 
   title        = "Quantum state transfer in arrays of flux qubits", 
   journal      = "New J. Phys.", 
   volume       = "7", 
   pages        = "181", 
   year         = "2005", 
   doi = "10.1088/1367-2630/7/1/181",
}

@ARTICLE{spin2a,
   author       = "Christandl, M. and Datta, N. and Ekert, A. and Landahl, A. J.", 
   title        = "Perfect state transfer in quantum spin networks", 
   journal      = "Phys. Rev. Lett.", 
   volume       = "92", 
   pages        = "187902", 
   year         = "2004", 
   doi = "10.1103/PhysRevLett.92.187902",
}

@ARTICLE{spin2b,
   author       = "Giovannetti, V. and Burgarth, D.", 
   title        = "Improved transfer of quantum information using a local memory", 
   journal      = "Phys. Rev. Lett.", 
   volume       = "96", 
   pages        = "030501", 
   year         = "2006", 
   doi = "10.1103/PhysRevLett.96.030501",
}

@ARTICLE{spin2c,
   author       = "Fitzsimons, J. and Twanley, J.", 
   title        = "Globally controlled quantum wires for perfect qubit transport, mirroring, and computing", 
   journal      = "Phys. Rev. Lett.", 
   volume       = "97", 
   pages        = "090502", 
   year         = "2006", 
   doi = "10.1103/PhysRevLett.97.090502",
}

@ARTICLE{spin2d,
   author       = "Burgarth, D.  and Bose,  S.", 
   title        = "Conclusive and arbitrarily perfect quantum-state transfer using parallel spin-chain channels", 
   journal      = "Phys. Rev. A", 
   volume       = "71", 
   pages        = "052315", 
   year         = "2005", 
   doi = "10.1103/PhysRevA.71.052315",
}

@ARTICLE{spin3,
   author       = "Wei, M. S. and Liao, M. J. and Wang, Y. Q. and Lin, Z. and Qiu, X. and Wang, C. and Xu, J. and Yang, Y.", 
   title        = "Quantum state transfer on square lattices with topology", 
   journal      = "Phys. Rev. A", 
   volume       = "108", 
   pages        = "062401", 
   year         = "2023", 
   doi = "10.1103/PhysRevA.108.062401",
}

@ARTICLE{ion1a,
   author       = "Parkins, A. S. and Kimble, H. J.", 
   title        = "Quantum state transfer between motion and light", 
   journal      = "J. Opt. B: Quantum Semiclass. Opt.", 
   volume       = "1", 
   pages        = "496", 
   year         = "1999", 
   doi = "10.1088/1464-4266/1/4/323",
}

@ARTICLE{ion1b,
   author       = "Nohama, F. K. and Roversi, J. A.", 
   title        = "Two-qubit state transfer between trapped ions using electromagnetic cavities coupled by an optical fibre", 
   journal      = "J. Phys. B: At. Mol. Opt. Phys.", 
   volume       = "41", 
   pages        = "045503", 
   year         = "2008", 
   doi = "10.1088/0953-4075/41/4/045503",
}

@ARTICLE{ion2a,
   author       = "Riebe, M. and H{\"{a}}ffner, H. and Roos, C. F. and H{\"{a}}nsel, W. and Benhelm, J. and Lancaster, G. P. T. and K{\"{o}}rber, T. W. and Becher, C. and  Schmidt-Kaler,F. and James, D. F. V. and Blatt, R.", 
   title        = "Deterministic quantum teleportation with atoms", 
   journal      = "Nature", 
   volume       = "429", 
   pages        = "734--737", 
   year         = "2004", 
   doi = "10.1038/nature02570",
}

@ARTICLE{ion2b,
   author       = "Barrett, M. D. and Chiaverini, J. and Schaetz, T. and Britton, J. and Itano, W. M. and Jost, J. D. and Knill, E. and Langer, C. and Leibfried, D. and Ozeri, R. and Wineland, D. J.", 
   title        = "Deterministic quantum teleportation of atomic qubits", 
   journal      = "Nature", 
   volume       = "429", 
   pages        = "737--739", 
   year         = "2004", 
   doi = "10.1038/nature02608",
}

@ARTICLE{ion2c,
   author       = "Riebe, M. and Chwalla, M. and Benhelm, J. and H{\"{a}}ffner, H. and H{\"{a}}nsel, W. and Roos, C. F. and Blatt, R.", 
   title        = "Quantum teleportation with atoms: quantum process tomography", 
   journal      = "New J. Phys.", 
   volume       = "9", 
   pages        = "211", 
   year         = "2007", 
   doi = "10.1088/1367-2630/9/7/211",
}

@ARTICLE{ion3,
   author       = "Olmschenk, S. and Matsukevich, D. N. and Maunz, P. and Hayes, D. and Duan, L.-M. and Monroe, C.", 
   title        = "Quantum teleportation between distant matter qubits", 
   journal      = "Science", 
   volume       = "323", 
   pages        = "486--489", 
   year         = "2009", 
   doi = "10.1126/science.1167209",
}

@ARTICLE{scc1a,
   author       = "Jahne, K. and Yurke, B. and Gavish, U.", 
   title        = "High-fidelity transfer of an arbitrary quantum state between harmonic oscillators", 
   journal      = "Phys. Rev. A", 
   volume       = "75", 
   pages        = "010301(R)", 
   year         = "2007", 
   doi = "10.1103/PhysRevA.75.010301",
}

@ARTICLE{scc1b,
   author       = "Srinivasan, S. J. and Sundaresan, N. M. and Sadri, D. and Liu, Y. and Gambetta, J. M. and Yu, T. and Girvin, S. M. and Houck, A. A.", 
   title        = "Time-reversal symmetrization of spontaneous emission for quantum state transfer", 
   journal      = "Phys. Rev. A", 
   volume       = "89", 
   pages        = "033857", 
   year         = "2014", 
   doi = "10.1103/PhysRevA.89.033857",
}

@ARTICLE{scc1c,
   author       = "Korotkov, A. N.", 
   title        = "Flying microwave qubits with nearly perfect transfer efficiency", 
   journal      = "Phys. Rev. B", 
   volume       = "84", 
   pages        = "014510", 
   year         = "2011", 
   doi = "10.1103/PhysRevB.84.014510",
}

@ARTICLE{scc1d,
   author       = "Sete, E. A. and Mlinar, E. and Korotkov, A. N.", 
   title        = "Robust quantum state transfer using tunable couplers", 
   journal      = "Phys. Rev. B", 
   volume       = "91", 
   pages        = "144509", 
   year         = "2015", 
   doi = "10.1103/PhysRevB.91.144509",
}

@ARTICLE{scc1e,
   author       = "Wenner, J. and Yin, J. and Chen, Y. and Barends, R. and Chiaro, B. and Jeffrey, E. and Kelly, J. and Megrant, A. and Mutus, J. Y. and Neill, C. and O'Malley, P. J. J. and Roushan, P. and Sank, D. and Vainsencher, A. and White, T. C. and Korotkov, A. N. and Cleland, A. N. and Martinis, J. M.", 
   title        = "Catching time-reversed microwave coherent state photons with 99.4\% absorption efficiency", 
   journal      = "Phys. Rev. Lett.", 
   volume       = "112", 
   pages        = "210501", 
   year         = "2014", 
   doi = "10.1103/PhysRevLett.112.210501",
}

@ARTICLE{scc1f,
   author       = "Axline, C. J. and Burkhart, L. D. and Pfaff, W. and Zhang, M. and Chou, K. and Campagne-Ibarcq, P. and Reinhold, P. and Frunzio, L. and Girvin, S. M. and Jiang, L. and Devoret, M. H. and Schoelkopf, R. J.", 
   title        = "On-demand quantum state transfer and entanglement between remote microwave cavity memories", 
   journal      = "Nature Phys.", 
   volume       = "14", 
   pages        = "705--710", 
   year         = "2018", 
   doi = "10.1038/s41567-018-0115-y",
}

@ARTICLE{scc1g,
   author       = "Kurpiers, P. and Magnard, P. and Walter, T. and Royer, B. and Pechal, M. and Heinsoo, J. and Salath\'{e}, Y. and Akin, A. and Storz, S. and Besse, J.-C. and Gasparinetti, S. and Blais, A. and Wallraff, A.", 
   title        = "Deterministic quantum state transfer and remote entanglement using microwave photons", 
   journal      = "Nature", 
   volume       = "558", 
   pages        = "264--267", 
   year         = "2018", 
   doi = "10.1038/s41586-018-0195-y",
}

@ARTICLE{scc1h,
   author       = "A. Bienfait and Satzinger, K. J. and Zhong, Y. P. and Chang, H.-S. and Chou, M.-H. and Conner, C. R. and Dumur, \'{E}. and Grebel, J. and Peairs, G. A. and Povey, R. G. and Cleland, A. N.", 
   title        = "Phonon-mediated quantum state transfer and remote qubit entanglement", 
   journal      = "Science", 
   volume       = "364", 
   pages        = "368--371", 
   year         = "2019", 
   doi = "10.1126/science.aaw8415",
}

@ARTICLE{scc2,
   author       = "Leung, N. and Lu, Y. and Chakram, S. and Naik, R. K. and Earnest, N. and Ma, R. and Jacobs, K. and Cleland, A. N. and Schuster, D. I.", 
   title        = "Deterministic bidirectional communication and remote entanglement generation between superconducting qubits", 
   journal      = "npj Quantum Inf.", 
   volume       = "5", 
   pages        = "18", 
   year         = "2019", 
   doi ="10.1038/s41534-019-0128-0",
}

@ARTICLE{scc3,
   author       = "Fedorov, K. G. and Renger, M. and Pogorzalek, S. and Candia, R. D. and Chen, Q. and Nojiri, Y. and Inomata, K. and Nakamura, Y. and Partanen, M. and Marx, A. and Gross, R. and Deppe, F.", 
   title        = "Experimental quantum teleportation of propagating microwaves", 
   journal      = "Sci. Adv.", 
   volume       = "7", 
   pages        = "eabk0891", 
   year         = "2021", 
   doi = "10.1126/sciadv.abk0891",
}

@ARTICLE{ss1,
   author       = "Lemonde, M.-A. and Meesala, S. and Sipahigil, A. and Schuetz, M. J. A. and Lukin, M. D. and Loncar, M. and Rabl, P.", 
   title        = "Phonon Networks with Silicon-Vacancy Centers in Diamond Waveguides", 
   journal      = "Phys. Rev. Lett.", 
   volume       = "120", 
   pages        = "213603", 
   year         = "2018", 
   doi ="10.1103/PhysRevLett.120.213603",
}

@ARTICLE{ss2a,
   author       = "Rozpedek, F. and Yehia, R. and Goodenough, K. and Ruf, M. and Humphreys, P. C. and Hanson, R. and Wehner, S. and Elkouss, D.", 
   title        = "Near-term quantum-repeater experiments with nitrogen-vacancy centers: {O}vercoming the limitations of direct transmission", 
   journal      = "Phys. Rev. A", 
   volume       = "99", 
   pages        = "052330", 
   year         = "2019", 
   doi = "10.1103/PhysRevA.99.052330",
}

@ARTICLE{ss2b,
   author       = "Pfaff, W. and Hensen, B. J. and Bernien, H. and Van Dam, S. B. and Blok, M. S. and Taminiau, T. H. and Tiggelman, M. J. and Schouten, R. N. and Markham, M. and Twitchen, D. J. and Hanson, R.", 
   title        = "Unconditional quantum teleportation between distant solid-state quantum bits", 
   journal      = "Science", 
   volume       = "345", 
   pages        = "532--535", 
   year         = "2014", 
   doi = "10.1126/science.1253512",
}

@ARTICLE{ss2c,
   author       = "M. Pompili and Hermans, S. L. N. and Baier, S. and Beukers, H. K. C. and Humphreys, P. C. and Schouten, R. N. and Vermeulen, R. F. L. and Tiggelman, M. J. and Dos Santos Martins, L. and Dirkse, B. and Wehner, S. and Hanson, R.", 
   title        = "Realization of a multinode quantum network of remote solid-state qubits", 
   journal      = "Science", 
   volume       = "372", 
   pages        = "259--264", 
   year         = "2021", 
   doi ="10.1126/science.abg1919",
}

@ARTICLE{opt1,
   author       = "Lin, Q. and He, B.", 
   title        = "Bi-directional mapping between polarization and spatially encoded photonic qutrits", 
   journal      = "Phys. Rev. A", 
   volume       = "80", 
   pages        = "062312", 
   year         = "2009", 
   doi = "10.1103/PhysRevA.80.062312",
}

@ARTICLE{opt2a,
   author       = "He, B. and Ren, Y. and Bergou, J. A.", 
   title        = "Creation of high-quality long-distance entanglement with flexible resources", 
   journal      = "Phys. Rev. A", 
   volume       = "79", 
   pages        = "052323", 
   year         = "2009", 
   doi = "10.1103/PhysRevA.79.052323",
}

@ARTICLE{opt2b,
   author       = "Jin, X.-M. and Ren, J.-G. and Yang, B. and Yi, Z.-H. and Zhou, F. and Xu, X.-F. and Wang, S.-K. and Yang, D. and Hu, Y.-F. and Jiang, S.", 
   title        = "Experimental free-space quantum teleportation", 
   journal      = "Nat. Photon.", 
   volume       = "4", 
   pages        = "376--381", 
   year         = "2010", 
   doi = "10.1038/nphoton.2010.87",
}

@ARTICLE{MTK,
   author       = "Aspelmeyer, M. and Kippenberg, T. J. and Marquardt, F.", 
   title        = "Cavity optomechanics", 
   journal      = "Rev. Mod. Phys.", 
   volume       = "86", 
   pages        = "1391", 
   year         = "2014", 
   doi = "10.1103/RevModPhys.86.1391",
}

@ARTICLE{SCExp1,
   author       = "Kleckner, D. and Bouwmeester, D.", 
   title        = "Sub-kelvin optical cooling of a micromechanical resonator", 
   journal      = "Nature", 
   volume       = "444", 
   pages        = "75--78", 
   year         = "2006", 
   doi = "10.1038/nature05231",
}

@ARTICLE{SCExp2,
   author       = "Gr{\"{o}}blacher, S. and Hammerer, K. and Vanner, M. R. and Aspelmeyer, M.", 
   title        = "Observation of strong coupling between a micromechanical resonator and an optical cavity field", 
   journal      = "Nature", 
   volume       = "460", 
   pages        = "724--727", 
   year         = "2009", 
   doi = "10.1038/nature08171",
}

@ARTICLE{SCExp3,
   author       = "O’Connell, A. D. and Hofheinz, M. and Ansmann, M. and Bialczak, R. C. and Lenander, M. and Lucero, E. and Neeley, M. and Sank, D. and Wang, H. and Weides, M. and Wenner, J. and Martinis, J. M. and Cleland, A. N.", 
   title        = "Quantum ground state and single-phonon control of a mechanical resonator", 
   journal      = "Nature", 
   volume       = "464", 
   pages        = "697--703", 
   year         = "2010",
   doi = "10.1038/nature08967", 
}

@ARTICLE{SCExp4,
   author       = "Teufel, J. D. and Donner, T. and Li, D. and Harlow,  J. W. and Allman, M. S. and Cicak, K. and Sirois, A. J. and Whittaker, J. D. and Lehnert, K. W. and Simmonds, R. W.", 
   title        = "Sideband cooling of micromechanical motion to the quantum ground state", 
   journal      = "Nature", 
   volume       = "475", 
   pages        = "359--363", 
   year         = "2011", 
   doi = "10.1038/nature10261",
}

@ARTICLE{SCExp5,
   author       = "Chan, J. and Mayer Alegre, T. P. and Safavi-Naeini, A. H. and Hill, J. T. and Krause, A. and Gr{\"{o}}blacher, S. and Aspelmeyer, M. and Painter, O.", 
   title        = "Laser cooling of a nanomechanical oscillator into its quantum ground state", 
   journal      = "Nature", 
   volume       = "478", 
   pages        = "89--92", 
   year         = "2011", 
   doi = "10.1038/nature10461",
}

@ARTICLE{SCExp6,
   author       = "Verhagen, E. and Del{\'{e}}glise, S. and Weis, S. and Schliesser, A. and Kippenberg, T. J.", 
   title        = "Quantum-coherent coupling of a mechanical oscillator to an optical cavity mode", 
   journal      = "Nature", 
   volume       = "482", 
   pages        = "63--67", 
   year         = "2012", 
   doi = "10.1038/nature10787",
}

@ARTICLE{CMExp1,
   author       = "Palomaki, T. A. and Harlow, J. W. and Teufel, J. D. and Simmonds, R. W. and Lehnert, K. W.", 
   title        = "Coherent state transfer between itinerant microwave fields and a mechanical oscillator", 
   journal      = "Nature", 
   volume       = "495", 
   pages        = "210--214", 
   year         = "2013", 
   doi ="10.1038/nature11915",
}

@ARTICLE{CMExp2,
   author       = "Fiore, V. and Yang, Y. and Kuzyk, M. C. and Barbour, R. and Tian, L. and Wang, H.", 
   title        = "Storing optical information as a mechanical excitation in a silica optomechanical resonator", 
   journal      = "Phys. Rev. Lett.", 
   volume       = "107", 
   pages        = "133601", 
   year         = "2011", 
   doi = "10.1103/PhysRevLett.107.133601",
}

@ARTICLE{CMExp3,
   author       = "Reed, A. P. and Mayer, K. H. and Teufel, J. D. and Burkhart, L. D. and Pfaff, W. and Reagor, M. and Sletten, L. and Ma, X. and Schoelkopf, R. J. and Knill, E. and Lehnert, K. W.", 
   title        = "Faithful conversion of propagating quantum information to mechanical motion", 
   journal      = "Nat. Phys.", 
   volume       = "13", 
   pages        = "1163--1167", 
   year         = "2017", 
   doi = "10.1038/nphys4251",
}

@ARTICLE{CMTheo1,
   author       = "Zhang, J. and Peng, K. and Braunstein, S. L.", 
   title        = "Quantum-state transfer from light to macroscopic oscillators", 
   journal      = "Phys. Rev. A", 
   volume       = "68", 
   pages        = "013808", 
   year         = "2003", 
   doi = "10.1103/PhysRevA.68.013808",
}

@ARTICLE{CMTheo2,
   author       = "J{\"{a}}hne, K. and Genes, C. and Hammerer, K. and Wallquist, M. and Polzik, E. S. and Zoller, P.", 
   title        = "Cavity-assisted squeezing of a mechanical oscillator", 
   journal      = "Phys. Rev. A", 
   volume       = "79", 
   pages        = "063819", 
   year         = "2009", 
   doi = "10.1103/PhysRevA.79.063819",
}

@ARTICLE{CMTheo3,
   author       = "Akram, U. and Kiesel, N. and Aspelmeyer, M. and Milburn, G. J.", 
   title        = "Single-photon opto-mechanics in the strong coupling regime", 
   journal      = "New J. Phys.", 
   volume       = "12", 
   pages        = "083030", 
   year         = "2010", 
   doi = "10.1088/1367-2630/12/8/083030",
}

@ARTICLE{CMTheo4,
   author       = "Khalili, F. and Danilishin, S. and Miao, H. and M{\"{u}}ller-Ebhardt, H. and Yang, H. and Chen, Y.", 
   title        = "Preparing a mechanical oscillator in non-{G}aussian quantum states", 
   journal      = "Phys. Rev. Lett.", 
   volume       = "105", 
   pages        = "070403", 
   year         = "2010", 
   doi = "10.1103/PhysRevLett.105.070403",
}

@ARTICLE{CMTheo5a,
   author       = "Teh, R. Y. and Kiesewetter, S. and Reid, M. D. and Drummond, P. D.", 
   title        = "Simulation of an optomechanical quantum memory in the nonlinear regime", 
   journal      = "Phys. Rev. A", 
   volume       = "96", 
   pages        = "013854", 
   year         = "2017", 
   doi = "10.1103/PhysRevA.96.013854",
}

@ARTICLE{CMTheo5b,
   author       = "Teh, R. Y. and Kiesewetter, S. and Drummond, P. D. and Reid, M. D.", 
   title        = "Creation, storage, and retrieval of an optomechanical cat state", 
   journal      = "Phys. Rev. A", 
   volume       = "98", 
   pages        = "063814", 
   year         = "2018", 
   doi = "10.1103/PhysRevA.98.063814",
}

@ARTICLE{CMTheo6,
   author       = "Cheng, J. and Liang, X.-T. and Zhang, W.-Z. and Duan, X.", 
   title        = "Optomechanical state transfer in the presence of non-{M}arkovian environments", 
   journal      = "Opt. Commun.", 
   volume       = "430", 
   pages        = "385--390", 
   year         = "2019", 
   doi = "10.1016/j.optcom.2018.08.079",
}

@ARTICLE{CMTheo7,
   author       = "Lei, S. and Wang, X. and Li, H. and Peng, R. and Xiong, B.", 
   title        = "High-fidelity and robust optomechanical state transfer based on pulse control", 
   journal      = "Appl. Phys. B", 
   volume       = "129", 
   pages        = "193", 
   year         = "2023", 
   doi = "10.1007/s00340-023-08135-3",
}

@ARTICLE{CMCTheo1,
   author       = "Tian, L. and Wang, H.", 
   title        = "Optical wavelength conversion of quantum states with optomechanics", 
   journal      = "Phys. Rev. A", 
   volume       = "82", 
   pages        = "053806", 
   year         = "2010", 
   doi = "10.1103/PhysRevA.82.053806",
}

@ARTICLE{CMCTheo2,
   author       = "Regal, C. A. and Lehnert, K. W.", 
   title        = "From cavity electromechanics to cavity optomechanics", 
   journal      = "J. Phys.: Conf. Ser.", 
   volume       = "264", 
   pages        = "012025", 
   year         = "2011", 
   doi = "10.1088/1742-6596/264/1/012025",
}

@ARTICLE{CMCTheo3,
   author       = "Tian, L.", 
   title        = "Adiabatic state conversion and pulse transmission in optomechanical systems", 
   journal      = "Phys. Rev. Lett.", 
   volume       = "108", 
   pages        = "153604", 
   year         = "2012", 
   doi = "10.1103/PhysRevLett.108.153604",
}

@ARTICLE{CMCTheo4,
   author       = "Wang, Y.-D. and Clerk, A. A.", 
   title        = "Using interference for high fidelity quantum state transfer in optomechanics", 
   journal      = "Phys. Rev. Lett.", 
   volume       = "108", 
   pages        = "153603", 
   year         = "2012", 
   doi = "10.1103/PhysRevLett.108.153603",
}

@ARTICLE{CMCTheo5,
   author       = "Tian, L.", 
   title        = "Optoelectromechanical transducer: Reversible conversion between microwave and optical photons", 
   journal      = "Ann. Phys. (Berlin)", 
   volume       = "527", 
   pages        = "1--14", 
   year         = "2015", 
   doi = "10.1002/andp.201400116",
}

@ARTICLE{CMCExp1,
   author       = "Hill, J. T. and Safavi-Naeini, A. H. and Chan, J. and Painter, O.", 
   title        = "Coherent optical wavelength conversion via cavity optomechanics", 
   journal      = "Nat. Commun.", 
   volume       = "3", 
   pages        = "1196", 
   year         = "2012", 
   doi = "10.1038/ncomms2201",
}

@ARTICLE{CMCExp2,
   author       = "Bochmann, J. and Vainsencher, A. and Awschalom, D. D. and Cleland, A. N.", 
   title        = "Nanomechanical coupling between microwave and optical photons", 
   journal      = "Nat. Phys.", 
   volume       = "9", 
   pages        = "712--716", 
   year         = "2013", 
   doi = "10.1038/nphys2748",
}

@ARTICLE{CMCExp3,
   author       = "Andrews, R. W. and Peterson, R. W. and Purdy, T. P. and Cicak,  K. and Simmonds, R. W. and Regal, C. A. and Lehnert, K. W.", 
   title        = "Bidirectional and efficient conversion between microwave and optical light", 
   journal      = "Nat. Phys.", 
   volume       = "10", 
   pages        = "321--326", 
   year         = "2014", 
   doi = "10.1038/nphys2911",
}

@ARTICLE{CMCExp4,
   author       = "Forsch, M. and Stockill, R. and Wallucks, A. and Marinkovi{\'{c}}, I. and G{\"{a}}rtner, C. and Norte, R. A. and van Otten, F. and Fiore, A. and Srinivasan, K. and Gr{\"{o}}blacher, S.", 
   title        = "Microwave-to-optics conversion using a mechanical oscillator in its quantum ground state", 
   journal      = "Nat. Phys.", 
   volume       = "16", 
   pages        = "69--74", 
   year         = "2020", 
   doi = "10.1038/s41567-019-0673-7",
}

@ARTICLE{HS1,
   author       = "Wang, Y.-D. and Clerk, A. A.", 
   title        = "Using dark modes for high-fidelity optomechanical quantum state transfer", 
   journal      = "New J. Phys.", 
   volume       = "14", 
   pages        = "105010", 
   year         = "2012", 
   doi = "10.1088/1367-2630/14/10/105010",
}

@ARTICLE{HS2,
   author       = "Sainadh U, S. and Narayanan, A.", 
   title        = "Mechanical switch for state transfer in dual-cavity optomechanical systems", 
   journal      = "Phys. Rev. A", 
   volume       = "88", 
   pages        = "033802", 
   year         = "2013", 
   doi = "10.1103/PhysRevA.88.033802",
}

@ARTICLE{HS3,
   author       = "McGee, S. A. and Meiser, D. and Regal, C. A. and Lehnert, K. W. and Holland, M. J.", 
   title        = "Mechanical resonators for storage and transfer of electrical and optical quantum states", 
   journal      = "Phys. Rev. A", 
   volume       = "87", 
   pages        = "053818", 
   year         = "2013", 
   doi = "10.1103/PhysRevA.87.053818",
}

@ARTICLE{StoA1,
   author       = "Gu{\'{e}}ry-Odelin, D. and Ruschhaupt, A. and Kiely, A. and Torrontegui, E. and Mart{\'{i}}nez-Garaot, S. and Muga, J. G.", 
   title        = "Shortcuts to adiabaticity: {C}oncepts, methods, and applications", 
   journal      = "Rev. Mod. Phys.", 
   volume       = "91", 
   pages        = "045001", 
   year         = "2019", 
   doi = "10.1103/RevModPhys.91.045001",
}

@ARTICLE{StoA3,
   author       = "Zhang, C.-L. and Chen, X. and Liao, C.-G. and Lin, X.-M.", 
   title        = "Reversible quantum state transfer in a three-mode optomechanical system", 
   journal      = "Laser Phys. Lett.", 
   volume       = "18", 
   pages        = "065206", 
   year         = "2021", 
   doi = "10.1088/1612-202X/abfe0d",
}

@ARTICLE{MCM1,
   author       = "Schmidt, M. and Ludwig, M. and Marquardt, F.", 
   title        = "Optomechanical circuits for nanomechanical continuous variable quantum state processing", 
   journal      = "New J. Phys.", 
   volume       = "14", 
   pages        = "125005", 
   year         = "2012", 
   doi = "10.1088/1367-2630/14/12/125005",
}

@ARTICLE{MCM2,
   author       = "Weaver, M. J. and Buters, F. and Luna, F. and Eerkens, H. and Heeck, K. and de Man, S. and Bouwmeester, D.", 
   title        = "Coherent optomechanical state transfer between disparate mechanical resonators", 
   journal      = "Nat. Commun.", 
   volume       = "8", 
   pages        = "824", 
   year         = "2017", 
   doi = "10.1038/s41467-017-00968-9",
}

@ARTICLE{MCM3,
   author       = "Zeng, Y.-X. and Shen, J. and Ding, M.-S. and Li, C.", 
   title        = "Macroscopic {S}chrödinger cat state swapping in optomechanical system", 
   journal      = "Opt. Express", 
   volume       = "28", 
   pages        = "9587--9602", 
   year         = "2020", 
   doi = "10.1364/OE.385814",
}

@ARTICLE{MCM4,
   author       = "Fedoseev, V. and Luna, F. and Hedgepeth, I. and L{\"{o}}ffler, W. and Bouwmeester, D.", 
   title        = "Stimulated {R}aman adiabatic passage in optomechanics", 
   journal      = "Phys. Rev. Lett.", 
   volume       = "126", 
   pages        = "113601", 
   year         = "2021", 
   doi = "10.1103/PhysRevLett.126.113601",
}

@ARTICLE{DisOMS1,
   author       = "Safavi-Naeini,A. H.  and Painter, O.", 
   title        = "Proposal for an optomechanical traveling wave phonon–photon translator", 
   journal      = "New J. Phys.", 
   volume       = "13", 
   pages        = "013017", 
   year         = "2011", 
   doi = "10.1088/1367-2630/13/1/013017",
}

@ARTICLE{DisOMS2,
   author       = "Fang, K. and Matheny, M. H. and Luan, X. and Painter, O.", 
   title        = "Optical transduction and routing of microwave phonons in cavity-optomechanical circuits", 
   journal      = "Nat. Photon.", 
   volume       = "10", 
   pages        = "489--496", 
   year         = "2016", 
   doi = "10.1038/nphoton.2016.107",
}

@ARTICLE{DisOMS3,
   author       = "de Moraes Neto, G. D. and Andrade, F. M. and Montenegro, V. and Bose, S.", 
   title        = "Quantum state transfer in optomechanical arrays", 
   journal      = "Phys. Rev. A", 
   volume       = "93", 
   pages        = "062339", 
   year         = "2016", 
   doi = "10.1103/PhysRevA.93.062339",
}

@ARTICLE{hybrid1a,
   author       = "Hammerer, K. and Wallquist, M. and Genes, C. and Ludwig, M. and Marquardt, F. and Treutlein, P. and Zoller, P. and Ye, J. and Kimble, H. J.", 
   title        = "Strong coupling of a mechanical oscillator and a single atom", 
   journal      = "Phys. Rev. Lett.", 
   volume       = "103", 
   pages        = "063005", 
   year         = "2009", 
   doi = "10.1103/PhysRevLett.103.063005",
}

@ARTICLE{hybrid1b,
   author       = "Wallquist, M. and Hammerer, K. and Zoller, P. and Genes, C. and Ludwig, M. and Marquardt, F. and Treutlein, P. and Ye, J. and Kimble, H. J.", 
   title        = "Single-atom cavity {QED} and optomicromechanics", 
   journal      = "Phys. Rev. A", 
   volume       = "81", 
   pages        = "023816", 
   year         = "2010", 
   doi = "10.1103/PhysRevA.81.023816",
}

@ARTICLE{hybrid2,
   author       = "Singh, S. and Jing, H. and Wright, E. M. and Meystre, P.", 
   title        = "Quantum-state transfer between a {B}ose-{E}instein condensate and an optomechanical mirror", 
   journal      = "Phys. Rev. A", 
   volume       = "86", 
   pages        = "021801(R)", 
   year         = "2012", 
   doi = "10.1103/PhysRevA.86.021801",
}

@ARTICLE{hybrid3,
   author       = "Momeni, F. and Naderi, M. H.", 
   title        = "Atomic quadrature squeezing and quantum state transfer in a hybrid atom–optomechanical cavity with two {D}uffing mechanical oscillators", 
   journal      = "J. Opt. Soc. Am. B", 
   volume       = "36", 
   pages        = "775--785", 
   year         = "2019", 
   doi = "10.1364/JOSAB.36.000775",
}

@ARTICLE{hybrid4,
   author       = "Molinares, H. and Eremeev, V. and Orszag, M.", 
   title        = "High-fidelity synchronization and transfer of quantum states in optomechanical hybrid systems", 
   journal      = "Phys. Rev. A", 
   volume       = "105", 
   pages        = "033708", 
   year         = "2022", 
   doi = "10.1103/PhysRevA.105.033708",
}

@ARTICLE{Ent,
   author       = "Sarma, A. K. and Chakraborty, S. and Kalita, S.", 
   title        = "Continuous variable quantum entanglement in optomechanical systems: {A} short review", 
   journal      = "AVS Quantum Sci.", 
   volume       = "3", 
   pages        = "015901", 
   year         = "2021", 
   doi = "10.1116/5.0022349",
}

@ARTICLE{Tel1,
   author       = "Pirandola, S.  and Mancini, S. and Vitali, D. and Tombesi, P.", 
   title        = "Continuous-variable entanglement and quantum-state teleportation between optical and macroscopic vibrational modes through radiation pressure", 
   journal      = "Phys. Rev. A", 
   volume       = "68", 
   pages        = "062317", 
   year         = "2003", 
   doi = "10.1103/PhysRevA.68.062317",
}

@ARTICLE{Tel2,
   author       = "Hofer, S. G. and Wieczorek, W. and Aspelmeyer, M. and Hammerer, K.", 
   title        = "Quantum entanglement and teleportation in pulsed cavity optomechanics", 
   journal      = "Phys. Rev. A", 
   volume       = "84", 
   pages        = "052327", 
   year         = "2011", 
   doi = "10.1103/PhysRevA.84.052327",
}

@ARTICLE{Tel3,
   author       = "Barzanjeh, Sh. and Abdi, M. and Milburn, G. J. and Tombesi, P. and Vitali, D.", 
   title        = "Reversible optical-to-microwave quantum interface", 
   journal      = "Phys. Rev. Lett.", 
   volume       = "109", 
   pages        = "130503", 
   year         = "2012", 
   doi = "10.1103/PhysRevLett.109.130503",
}

@ARTICLE{Tel4,
   author       = "Pautrel, S. and Denis, Z. and Bon, J. and Borne, A. and Favero, I.", 
   title        = "Optomechanical discrete-variable quantum teleportation scheme", 
   journal      = "Phys. Rev. A", 
   volume       = "101", 
   pages        = "063820", 
   year         = "2020", 
   doi = "10.1103/PhysRevA.101.063820",
}

@ARTICLE{Tel5,
   author       = "Li, J. and Wallucks, A. and Benevides, R. and Fiaschi, N. and Hensen, B. and Mayer Alegre, T. P. and Gr{\"{o}}blacher, S.", 
   title        = "Proposal for optomechanical quantum teleportation", 
   journal      = "Phys. Rev. A", 
   volume       = "102", 
   pages        = "032402", 
   year         = "2020", 
   doi = "10.1103/PhysRevA.102.032402",
}

@ARTICLE{Tel6,
   author       = "Fiaschi, N. and Hensen, B. and Wallucks, A. and Benevides, R. and Li, J. and Mayer Alegre, T. P. and Gr{\"{o}}blacher, S.", 
   title        = "Optomechanical quantum teleportation", 
   journal      = "Nat. Photon.", 
   volume       = "15", 
   pages        = "817--821", 
   year         = "2021", 
   doi = "10.1038/s41566-021-00866-z",
}

@ARTICLE{Law,
   author       = "Law, C. K.", 
   title        = "Interaction between a moving mirror and radiation pressure: {A} {H}amiltonian formulation", 
   journal      = "Phys. Rev. A", 
   volume       = "51", 
   pages        = "2537", 
   year         = "1995", 
   doi = "10.1103/PhysRevA.51.2537",
}

@ARTICLE{JLim,
   author       = "Lim, J. and Tame, M. and Yee, K. H. and Lee, J.-S. and Lee, J.", 
   title        = "Phonon-induced dynamic resonance energy transfer", 
   journal      = "New J. Phys.", 
   volume       = "16", 
   pages        = "053018", 
   year         = "2014", 
   doi = "10.1088/1367-2630/16/5/053018",
}

@BOOK{Knight,
   author       = {Gerry, C. C. and Knight, P. L.},
   year         = {2005},
   title        = {Introductory quantum optics},
   publisher    = {Cambridge University Press},
   doi = {10.1017/CBO9780511791239},
}

@ARTICLE{JLee,
   author       = "Lee, C.-W. and Lee, J. and Nha, H. and Jeong, H.", 
   title        = "Generating a {S}chrödinger-cat-like state via a coherent superposition of photonic operations", 
   journal      = "Phys. Rev. A", 
   volume       = "85", 
   pages        = "063815", 
   year         = "2012", 
   doi = "10.1103/PhysRevA.85.063815",
}

@ARTICLE{SCS1,
   author       = "Vlastakis, B. and Kirchmair, G. and Leghtas, Z. and Nigg, S. E. and Frunzio, L. and Girvin, S. M. and Mirrahimi, M. and Devoret, M. H. and Schoelkopf, R. J.", 
   title        = "Deterministically encoding quantum information using 100-photon {S}chrödinger cat states", 
   journal      = "Science", 
   volume       = "342", 
   pages        = "607--610", 
   year         = "2013", 
   doi = "10.1126/science.1243289",
}

@ARTICLE{SCS2,
   author       = "Liao, J.-Q. and Tian, L.", 
   title        = "Macroscopic quantum superposition in cavity optomechanics", 
   journal      = "Phys. Rev. Lett.", 
   volume       = "116", 
   pages        = "163602", 
   year         = "2016", 
   doi = "10.1103/PhysRevLett.116.163602",
}

@ARTICLE{SCS3,
   author       = "Hoff, U. B. and Kollath-B{\"{o}}nig, J. and Neergaard-Nielsen, J. S. and Andersen, U. L.", 
   title        = "Measurement-Induced Macroscopic Superposition States in Cavity Optomechanics", 
   journal      = "Phys. Rev. Lett.", 
   volume       = "117", 
   pages        = "143601", 
   year         = "2016", 
   doi = "10.1103/PhysRevLett.117.143601",
}

@BOOK{GZ04,
   author       = {Gardiner, C. and Zoller,  P.},
   year         = 2004,
   title        = {Quantum Noise},
   publisher    = {Springer Berlin, Heidelberg},
   doi=false,}

@ARTICLE{CROW,
   author       = "Yariv, A. and Xu,  Y. and Lee,  R. K. and Scherer, A.", 
   title        = "Coupled-resonator optical waveguide: {A} proposal and analysis", 
   journal      = "Opt. Lett.", 
   volume       = "24", 
   pages        = "711--713", 
   year         = "1999", 
   doi = "10.1364/OL.24.000711",
}

@ARTICLE{Lud,
   author       = "Ludwig, M. and Kubala, B. and Marquardt, F.", 
   title        = "The optomechanical instability in the quantum regime", 
   journal      = "New J. Phys.", 
   volume       = "10", 
   pages        = "095013", 
   year         = "2008", 
   doi = "10.1088/1367-2630/10/9/095013",
}

@ARTICLE{Murch,
   author       = "Murch, K. W. and Moore, K. L. and Gupta, S. and Stamper-Kurn, D. M.", 
   title        = "Observation of quantum-measurement backaction with an ultracold atomic gas", 
   journal      = "Nat. Phys.", 
   volume       = "4", 
   pages        = "561--564", 
   year         = "2008",
   doi = "10.1038/nphys965", 
}

@ARTICLE{Burgwal,
   author       = "Burgwal, R. and Verhagen,  E.", 
   title        = "Enhanced nonlinear optomechanics in a coupled-mode photonic crystal device", 
   journal      = "Nat. Commun.", 
   volume       = "14", 
   pages        = "1526", 
   year         = "2023", 
   doi = "10.1038/s41467-023-37138-z",
}

@ARTICLE{PhaseShifter1,
   author       = "Smith, B. J. and Kundys, D. and Thomas-Peter, N. and Smith, P. G. R. and Walmsley, I. A.", 
   title        = "Phase-controlled integrated photonic quantum circuits", 
   journal      = "Opt. Express", 
   volume       = "17", 
   pages        = "13516--13525", 
   year         = "2009", 
   doi = "10.1364/OE.17.013516",
}

@ARTICLE{PhaseShifter2,
   author       = "Kokkoniemi, R. and Ollikainen, T. and Lake, R. E. and Saarenp{\"{a}}{\"{a}}, S. and Tan, K. Y. and Kokkala, J. I. and Da{\u{g}}, C. B. and Govenius, J. and M{\"{o}}tt{\"{o}}nen, M.", 
   title        = "Flux-tunable phase shifter for microwaves", 
   journal      = "Sci. Rep.", 
   volume       = "7", 
   pages        = "14713", 
   year         = "2017", 
   doi = "10.1038/s41598-017-15190-2",
}

@ARTICLE{Adia2,
   author       = "Sarandy, M. S. and Lidar, D. A.", 
   title        = "Adiabatic approximation in open quantum systems", 
   journal      = "Phys. Rev. A", 
   volume       = "71", 
   pages        = "012331", 
   year         = "2005", 
   doi = "10.1103/PhysRevA.71.012331",
}

@ARTICLE{CSstate,
   author       = "M{\o{}}ller, K. B. and J{\o{}}rgensen, T. G. and Dahl, J. P.", 
   title        = "Displaced squeezed number states: {P}osition space representation, inner product, and some applications", 
   journal      = "Phys. Rev. A", 
   volume       = "54", 
   pages        = "5378", 
   year         = "1996", 
   doi = "10.1103/PhysRevA.54.5378",
}

@ARTICLE{Lemonde,
   author       = "Lemonde, M.-A. and Didier, N. and Clerk, A.", 
   title        = "Enhanced nonlinear interactions in quantum optomechanics via mechanical amplification", 
   journal      = "Nat. Commun.", 
   volume       = "7", 
   pages        = "11338", 
   year         = "2016", 
   doi = "10.1038/ncomms11338",
}

@ARTICLE{Manninen,
   author       = "Manninen, J. and Haque, M. T. and Vitali, D. and Hakonen, P.", 
   title        = "Enhancement of the optomechanical coupling and {K}err nonlinearity using the {J}osephson capacitance of a {C}ooper-pair box", 
   journal      = "Phys. Rev. B", 
   volume       = "105", 
   pages        = "144508", 
   year         = "2022", 
   doi = "10.1103/PhysRevB.105.144508",
}

@ARTICLE{Braunstein,
   author       = "Braunstein, S. L. and van Loock,  P.", 
   title        = "Quantum information with continuous variables", 
   journal      = "Rev. Mod. Phys.", 
   volume       = "77", 
   pages        = "513", 
   year         = "2005", 
   doi = "10.1103/RevModPhys.77.513",
}

@ARTICLE{Fukui,
   author       = "Fukui, K. and Takeda, S.", 
   title        = "Building a large-scale quantum computer with continuous-variable optical technologies", 
   journal      = "J. Phys. B: At. Mol. Opt. Phys.", 
   volume       = "55", 
   pages        = "012001", 
   year         = "2022", 
   doi = "10.1088/1361-6455/ac489c",
}

@ARTICLE{mem1,
   author       = "Thompson, J. D. and Zwickl, B. M. and Jayich, A. M. and Marquardt, F. and Girvin, S. M. and Harris, J. G. E.", 
   title        = "Strong dispersive coupling of a high-finesse cavity to
a micromechanical membrane", 
   journal      = "Nature", 
   volume       = "452", 
   pages        = "72", 
   year         = "2008", 
   doi = "10.1038/nature06715"
}

@ARTICLE{mem2,
   author       = "Biancofiore, C. and Karuza, M. and Galassi, M. and Natali, R. and Tombesi, P. and Di Giuseppe, G. and Vitali, D.", 
   title        = "Quantum dynamics of an optical cavity coupled to a thin semitransparent membrane:
{E}ffect of membrane absorption", 
   journal      = "Phys. Rev. A", 
   volume       = "84", 
   pages        = "033814", 
   year         = "2011", 
   doi = "10.1103/PhysRevA.84.033814"
}

@ARTICLE{add2,
   author       = "Dong, C. and Fiore, V. and Kuzyk, M. C. and Wang, H.", 
   title        = "Optomechanical Dark Mode", 
   journal      = "Science", 
   volume       = "338", 
   pages        = "1609--1613", 
   year         = "2012", 
   doi = "10.1126/science.1228370"
}

@ARTICLE{add4,
   author       = "Kues, M. and Reimer, C. and Lukens, J. M. and Munro, W. J. and Weiner, A. M. and Moss, D. J. and Morandotti, R.", 
   title        = "Quantum optical microcombs", 
   journal      = "Nat. Photon.", 
   volume       = "13", 
   pages        = "170--179", 
   year         = "2019", 
   doi = "10.1038/s41566-019-0363-0"
}

@ARTICLE{add5,
   author       = "Labont{\'{e}}, L. and Alibart, O. and D’Auria, V. and Doutre, F. and Etesse, J. and Sauder, G. and Martin, A. and Picholle, {\'{E}}. and Tanzilli, S.", 
   title        = "Integrated Photonics for Quantum Communications and Metrology", 
   journal      = "PRX Quantum", 
   volume       = "5", 
   pages        = "010101", 
   year         = "2024", 
   doi = "10.1103/PRXQuantum.5.010101"
}

@ARTICLE{add6,
   author       = "Fogliano, F. and Besga, B. and Reigue, A. and Heringlake, P. and de L{\'{e}}pinay, L. M. and Vaneph, C. and Reichel, J. and Pigeau, B. and Arcizet, O.", 
   title        = "Mapping the Cavity Optomechanical Interaction with Subwavelength-Sized Ultrasensitive Nanomechanical Force Sensors", 
   journal      = "Phys. Rev. X", 
   volume       = "11", 
   pages        = "021009", 
   year         = "2021", 
   doi = "10.1103/PhysRevX.11.021009"
}

@ARTICLE{add7,
   author       = "Berry, M. V.", 
   title        = "Transitionless quantum driving", 
   journal      = "J. Phys. A: Math. Theor.", 
   volume       = "42", 
   pages        = "365303", 
   year         = "2009", 
   doi = "10.1088/1751-8113/42/36/365303"
}

@ARTICLE{add8,
   author       = "Sels, D. and Polkovnikov, A.", 
   title        = "Minimizing irreversible losses in quantum systems by local counterdiabatic driving", 
   journal      = "Proc. Natl Acad. Sci. USA", 
   volume       = "114", 
   pages        = "E3909--E3916", 
   year         = "2017", 
   doi = "10.1073/pnas.1619826114"
}

@ARTICLE{add9,
   author       = "Weigel, P. O.  and Zhao, J.  and Fang, K. and Al-Rubaye, H. and Trotter, D. and Hood, D. and Mudrick, J. and Dallo, C. and Pomerene, A. T. and Starbuck, A. L. and DeRose, C. T. and Lentine, A. L. and Rebeiz, G. and Mookherjea,  S.", 
   title        = "Bonded thin film lithium niobate modulator on a silicon photonics platform exceeding 100 {GH}z 3-d{B} electrical modulation bandwidth", 
   journal      = "Opt. Express", 
   volume       = "26", 
   pages        = "23728--23739", 
   year         = "2018", 
   doi = "10.1364/OE.26.023728"
}

@ARTICLE{add10,
   author       = "Barbiero, M. and Salvatierra, J. P. and Risaro, M.  and Clivati, C. and Calonico, D. and Levi, F. and Tarallo, M. G.", 
   title        = "Broadband serrodyne phase modulation for optical frequency standards and spectral purity transfer", 
   journal      = "Opt. Lett.", 
   volume       = "48", 
   pages        = "1958--1961", 
   year         = "2023", 
   doi = "10.1364/OL.485064"
}

@ARTICLE{add11,
   author       = "Rossi, M. and Mason, D. and Chen, J. and Tsaturyan, Y. and Schliesser, A.", 
   title        = "Measurement-based quantum control of mechanical motion", 
   journal      = "Nature", 
   volume       = "563", 
   pages        = "53--58", 
   year         = "2018", 
   doi = "10.1038/s41586-018-0643-8"
}

@ARTICLE{add12,
   author       = "Liao, J.-Q. and Huang, J.-F. and Tian, L. and Kuang, L.-M. and Sun, C.-P.", 
   title        = "Generalized ultrastrong optomechanical-like coupling", 
   journal      = "Phys. Rev. A", 
   volume       = "101", 
   pages        = "063802", 
   year         = "2020", 
   doi = "10.1103/PhysRevA.101.063802"
}

@ARTICLE{add13,
   author       = "Yin, X.-L. and Zhou, Y.-H. and Liao, J.-Q.", 
   title        = "All-optical quantum simulation of ultrastrong optomechanics", 
   journal      = "Phys. Rev. A", 
   volume       = "105", 
   pages        = "013504", 
   year         = "2022", 
   doi = "10.1103/PhysRevA.105.013504"
}

@ARTICLE{add14,
   author       = "Qin, W. and Kockum, A. F. and Mu{\~{n}}oz, C.  S and Miranowicz, A. and Nori, F.", 
   title        = "Quantum amplification and simulation of strong and ultrastrong coupling of light and matter", 
   journal      = "Phys. Rep.", 
   volume       = "1078", 
   pages        = "1--59", 
   year         = "2024", 
   doi = "10.1016/j.physrep.2024.05.003"
}

@ARTICLE{add15,
   author       = "An, S. and Lv, D. and del Campo, A. and Kim, K.", 
   title        = "Shortcuts to adiabaticity by counterdiabatic
driving for trapped-ion displacement in phase
space", 
   journal      = "Nat. Commun.", 
   volume       = "7", 
   pages        = "12999", 
   year         = "2016", 
   doi = "10.1038/ncomms12999"
}

@ARTICLE{add16,
   author       = "Kounalakis, M. and Blanter, Y. M. and Steele, G. A.", 
   title        = "Flux-mediated optomechanics with a transmon qubit in the single-photon
ultrastrong-coupling regime", 
   journal      = "Phys. Rev. Res.", 
   volume       = "2", 
   pages        = "023335", 
   year         = "2020", 
   doi = "10.1103/PhysRevResearch.2.023335"
}

\end{document}